\ifdefined\PABReview
  \let\PABPreprint\undefined
\else
  \def\PABPreprint{1}
\fi
\documentclass{article}
\usepackage[T1]{fontenc}
\usepackage{microtype}
\usepackage{amsmath}
\usepackage{graphicx,booktabs,array,tabularx,xcolor,colortbl,pifont,wasysym}
\usepackage{float}
\usepackage{hyperref}
\IfFileExists{latexml.sty}{\usepackage{latexml}}{\newif\iflatexml\latexmlfalse}
\ifdefined\PABPreprint
  \usepackage[preprint]{icml2027}
\else
  \usepackage{icml2027}
\fi
\ifdefined\PABPreprint
  \hypersetup{pdfsubject={Preprint}}
\else
  \hypersetup{pdfsubject={Anonymous manuscript under review}}
\fi
\usepackage{amsmath,amsfonts,bm}

\def\eqref#1{equation~\ref{#1}}
\def\1{\bm{1}}

\DeclareMathAlphabet{\mathsfit}{\encodingdefault}{\sfdefault}{m}{sl}
\SetMathAlphabet{\mathsfit}{bold}{\encodingdefault}{\sfdefault}{bx}{n}

\usepackage{amssymb}
\usepackage{enumitem}
\DeclareMathSizes{7.4}{7.4}{5}{5}
\AtEndDocument{%
  \ifdefined\PABPreprint\typeout{PAB-MODE=PREPRINT}\else\typeout{PAB-MODE=REVIEW}\fi
  \typeout{PAB-MAIN-LAST-PAGE=\getpagerefnumber{pab:main-end}}%
  \typeout{PAB-STATEMENTS-FIRST-PAGE=\getpagerefnumber{pab:statements-start}}%
}
\icmltitlerunning{ProactiveAudioBench: Condition-Driven Responses in Audio Streams}
\begin{document}
\twocolumn[
\icmltitle{ProactiveAudioBench: Evaluating Condition-Driven Responses in Audio Streams}
% Author order and affiliations retained from the existing named-author manuscript.
\begin{icmlauthorlist}
  \icmlauthor{Meng Gao}{thu,ath}
  \icmlauthor{Yunfei Chu}{ath}
  \icmlauthor{Yifan Yang}{sjtu,ath}
  \icmlauthor{Hui Wang}{nankai,ath}
  \icmlauthor{Wenxiang Guo}{zju,ath}
  \icmlauthor{Qi Chen}{sjtu,ath}
  \icmlauthor{Zihan Liu}{ath}
  \icmlauthor{Ling Wang}{polyu,ath}
  \icmlauthor{Yinsong Yan}{polyu,ath}
  \icmlauthor{Wen Huang}{jhu,ath}
  \icmlauthor{Xize Cheng}{ath}
  \icmlauthor{Yuxuan Wang}{ath}
  \icmlauthor{Jin Xu}{ath}
  \icmlauthor{Yujiu Yang}{thu}
\end{icmlauthorlist}
\icmlaffiliation{thu}{Tsinghua University}
\icmlaffiliation{ath}{Alibaba Token Hub, Alibaba Group}
\icmlaffiliation{sjtu}{Shanghai Jiao Tong University}
\icmlaffiliation{nankai}{Nankai University}
\icmlaffiliation{zju}{Zhejiang University}
\icmlaffiliation{polyu}{Hong Kong Polytechnic University}
\icmlaffiliation{jhu}{Johns Hopkins University}
% Corresponding authors remain Jin Xu and Yujiu Yang, as in the confirmed manuscript.
% User-provided author email addresses (2026-11-29); no email was supplied for Jin Xu.
\ifdefined\PABPreprint
  \icmlcorrespondingauthor{Jin Xu and Yujiu Yang}{yang.yujiu@sz.tsinghua.edu.cn}
  \gdef\PABContactDetails{%
    Jin Xu and Yujiu Yang \mbox{\textless{}\href{mailto:yang.yujiu@sz.tsinghua.edu.cn}{\nolinkurl{yang.yujiu@sz.tsinghua.edu.cn}}\textgreater{}}.
    Author email: Meng Gao \mbox{\textless{}\href{mailto:gaom24@mails.tsinghua.edu.cn}{\nolinkurl{gaom24@mails.tsinghua.edu.cn}}\textgreater{}}%
  }
  \makeatletter
  \gdef\icmlcorrespondingauthor@text{\PABContactDetails}
  \makeatother
\else
  \icmlcorrespondingauthor{Anonymous Author}{}
\fi

\icmlkeywords{Audio understanding, proactive interaction, streaming evaluation, benchmarks}
\vskip 0.3in
]
\printAffiliationsAndNotice{}
% The ICML HTML binding omits the template's affiliation/contact footnote.
\iflatexml
  \ifdefined\PABPreprint
    {\small\noindent Correspondence to: \PABContactDetails.\par}
  \fi
\fi
\begin{abstract}
As multimodal models advance from offline understanding to real-time interaction, audio assistants must continuously monitor user-specified targets and respond proactively when the requested conditions are met. Existing evaluations, however, do not systematically cover the diverse listening targets and notification requirements across sound, speech, and music. We introduce \textbf{ProactiveAudioBench}, comprising four task families and 22 subtasks across these three domains, including pronunciation correction, tool calling, musical section tracking, and matching against reference audio. These tasks combine diverse listening targets with rules specifying notifications at the first, every, or $N$th qualifying occurrence, while also testing whether models remain silent when conditions are unmet. The benchmark contains 1,999 listening requests with reference notification times and responses, curated through automated checks and two-stage human review. Our evaluation framework supports both simulated and native streaming and jointly assesses notification timing and response content. Experiments across 23 model–interface configurations show that high event coverage does not ensure reliable execution of listening requests, as models struggle to avoid both omissions and extra outputs. Further diagnostics reveal limitations in event counting and temporal grounding even without autonomous notification timing, suggesting that reliable listening requires improvements in both rule execution and the underlying audio capabilities.

\end{abstract}
\begin{figure*}[t]
\centering
\includegraphics[width=\linewidth]{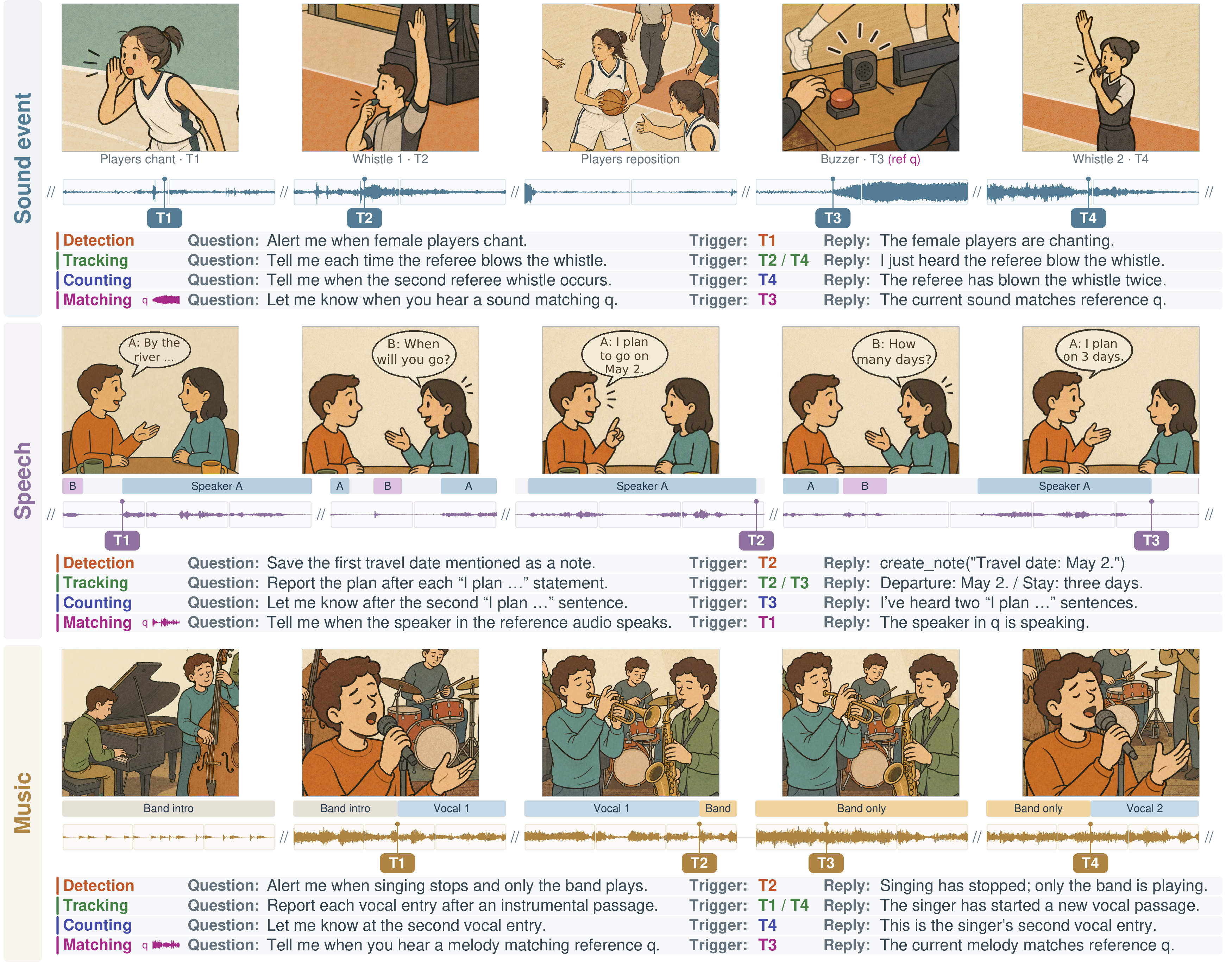}
\caption{Overview of ProactiveAudioBench.}
\label{fig:main-examples}
\end{figure*}

\section{Introduction}
\label{sec:introduction}

Multimodal large language models are moving from understanding and answering questions about fixed inputs toward continuous perception and real-time interaction. Benchmarks such as MMAU~\citep{arxiv2410_19168} and MMAR~\citep{arxiv2505_13032} evaluate understanding and reasoning across sound, speech, and music. In continuous interaction, audio assistants must also follow listening requests, autonomously deciding whether to respond, when to respond, and what to say as audio arrives.

Recent work has begun to evaluate these proactive interactions. Benchmarks such as OmniPro~\citep{arxiv2605_18577} and OmniInteract~\citep{arxiv2605_26485} examine monitoring, counting, and continuous responses in audiovisual streams. In audio, ProVoice-Bench~\citep{arxiv2604_15037} studies proactive intervention and tool actions, while Proactive-Sound-Bench~\citep{arxiv2606_05121} evaluates responses and silence under predefined event policies.

Existing evaluations provide limited coverage of how diverse listening targets and user-specified notification rules jointly govern responses over time. A request may target a sound category, a pronunciation error, a musical section, or a speaker specified through reference audio. For recurring targets, first-, every-, and $N$th-occurrence instructions require different notification sequences. Evaluating continuous listening therefore requires assessing the resulting response sequence, including both required notifications and periods of silence.

We introduce \textbf{ProactiveAudioBench} to evaluate the execution of user-specified listening conditions across sound events, speech, and music. \textbf{Detection}, \textbf{Tracking}, and \textbf{Counting} cover first-, every-, and specified $N$th-occurrence notification rules, respectively. Alongside these rules, \textbf{Matching} evaluates targets specified through reference audio and requires one notification at the first match. These four task families encompass 22 subtasks, including pronunciation correction, tool calling, and musical section tracking.

We combine carefully designed audio annotation and controlled audio construction with automated checks and two-stage human review to produce 1,999 listening requests with reference notification times and replies. The benchmark includes cases requiring silence when the listening condition is not satisfied. Our evaluation framework supports simulated and native streaming, jointly assessing notification timing and response content.

Evaluation across 23 model--interface configurations shows that high event coverage does not guarantee complete execution of listening requests, as models struggle to avoid both omissions and extra outputs. Counting and interval-localization probes further reveal difficulties in accurately counting and locating audio events even when models need not choose when to notify.

Our main contributions are:
\begin{itemize}[leftmargin=*, itemsep=0pt, topsep=2pt]
\item \textbf{A benchmark for sustained execution of listening requests.} We organize four task families and 22 subtasks across sound, speech, and music around target specification and notification rules, with human-reviewed notification times and reference replies.
\item \textbf{Streaming evaluation of notification timing and content.} Our framework supports simulated and native streaming, accounting for valid notifications, omissions, and extra outputs to evaluate sustained execution of listening requests.
\item \textbf{Systematic analysis of execution reliability and underlying capabilities.} We evaluate different models and interfaces, combining response-behavior analysis with counting and localization probes to examine execution difficulties and limitations in basic audio capabilities.
\end{itemize}

\section{Related Work}
\label{sec:related-work}

\subsection{Audio Understanding and Interactive Evaluation}
\label{sec:related-understanding}

Audio benchmarks assess complementary capabilities needed for continuous assistance. MMAU and MMAR evaluate understanding and reasoning across sound, speech, and music~\citep{arxiv2410_19168,arxiv2505_13032}, while MECAT targets fine-grained audio captioning and question answering~\citep{mecat2026}. CMI-Bench and BASS examine musical attributes, structure, and semantics~\citep{arxiv2506_12285,arxiv2602_04085}. Temporal audio grounding and reference-conditioned sound detection further assess when a specified target occurs~\citep{arxiv2609_01542,arxiv2112_10153,arxiv2507_16343}. These evaluations assess target understanding, while continuous interaction additionally requires deciding whether and when to respond as evidence arrives.

Full-duplex evaluation addresses this interaction dimension. Full-Duplex-Bench measures conversational turn-taking~\citep{arxiv2503_04721}, and DuplexAct-Bench extends behavioral coverage to proactive initiation and active silence under different instruction conditions~\citep{arxiv2609_39446}. $\tau$-Voice combines full-duplex conversation with verifiable task completion~\citep{tauvoice2026}. These benchmarks emphasize conversational participation and task-oriented dialogue, complementing evaluation of persistent audio-listening requests.

\subsection{Proactive Monitoring and Response Evaluation}
\label{sec:related-interaction}

% Requires booktabs, array, graphicx, xcolor, colortbl, pifont, wasysym, natbib.
\begin{table*}[t]
\centering
\caption{Comparison of ProactiveAudioBench with existing proactive interaction benchmarks.}
\label{tab:benchmark-comparison}

\begingroup
\definecolor{cmpgreen}{HTML}{217A57}
\definecolor{cmpred}{HTML}{AC4646}
\definecolor{cmpamber}{HTML}{B07822}
\definecolor{cmplight}{HTML}{ECF3F7}

\newcommand{\cmpyes}{\textcolor{cmpgreen}{\ding{51}}}
\newcommand{\cmpno}{\textcolor{cmpred}{\ding{55}}}
\newcommand{\cmppart}{\textcolor{cmpamber}{\LEFTcircle}}

% Column types: vertically centered and consistently aligned.
\newcolumntype{L}[1]{>{\raggedright\arraybackslash}m{#1}}
\newcolumntype{C}[1]{>{\centering\arraybackslash}m{#1}}

\setlength{\tabcolsep}{2.5pt}
\renewcommand{\arraystretch}{1.15}

\resizebox{\linewidth}{!}{%
\fontsize{8}{9.4}\selectfont
\begin{tabular}{@{}L{118pt}*{7}{C{40pt}}@{}}
\toprule
&
\multicolumn{3}{c}{\textbf{Domains}} &
\multicolumn{4}{c}{\textbf{Proactive Listening Tasks}} \\
\cmidrule(lr){2-4}
\cmidrule(lr){5-8}

\textbf{Benchmark}
& \textbf{Sound}
& \textbf{Speech}
& \textbf{Music}
& \shortstack{\textbf{Detection}\\\scriptsize First}
& \shortstack{\textbf{Tracking}\\\scriptsize Every}
& \shortstack{\textbf{Counting}\\\scriptsize Nth-only}
& \shortstack{\textbf{Matching}\\\scriptsize Acoustic ref.} \\
\midrule

ProVoice-Bench
& \cmpyes & \cmpyes & \cmpno
& \cmppart & \cmpno & \cmpno & \cmpno \\

Proactive-Sound-Bench (AIM)
& \cmpyes & \cmpno & \cmppart
& \cmppart & \cmppart & \cmpno & \cmpno \\

OmniPro
& \cmpyes & \cmpyes & \cmppart
& \cmpyes & \cmpyes & \cmppart & \cmpno \\

OmniInteract
& \cmppart & \cmpyes & \cmpno
& \cmpyes & \cmppart & \cmpno & \cmpno \\

\midrule
\rowcolor{cmplight}
\textbf{ProactiveAudioBench}~{\scriptsize(ours)}
& \cmpyes & \cmpyes & \cmpyes
& \cmpyes & \cmpyes & \cmpyes & \cmpyes \\
\bottomrule
\end{tabular}%
}

\vspace{3pt}
\begin{minipage}{\linewidth}
\fontsize{7}{8.2}\selectfont
\cmpyes\ explicit coverage;
\cmppart\ related or partial coverage;
\cmpno\ not evaluated.
OmniPro and OmniInteract are audiovisual benchmarks; OmniPro is compared in its online setting.
\end{minipage}

\endgroup
\end{table*}

Proactive video and audiovisual benchmarks evaluate responses tied to evolving perceptual evidence. ProactiveVideoQA and PhoStream examine answer timing as evidence becomes available during video or audiovisual streaming~\citep{arxiv2507_09313,phostream2026}. OmniPro covers alerts, monitoring, grounding, counting, narration, and procedural guidance in audiovisual streams~\citep{arxiv2605_18577}. OmniInteract evaluates continuous audiovisual monitoring, interruptions, and task resumption~\citep{arxiv2605_26485}, while StreamArena extends evaluation to hour-scale audiovisual interaction and memory~\citep{arxiv2608_05703}. DIVE further combines proactive intervention, continuous tracking, and interaction coordination, separately assessing participation, timing, and response quality~\citep{arxiv2610_28405}. These benchmarks establish important streaming interaction requirements, with visual or audiovisual evidence central to their tasks.

Audio-only proactive assistance has also been studied directly. \emph{I'll Keep an Ear Out} trains AudioLLMs to monitor a single natural-language intent, detecting relevant sound onsets and suppressing irrelevant or duplicate alerts~\citep{arxiv2609_21183}. ProVoice-Bench combines conversational audio with digital context to evaluate intent capture, topic monitoring, fact checking, and environmental sound sensing, including user-defined triggers~\citep{arxiv2604_15037}. Its reported evaluation assesses intervention decisions and the correctness of responses and tool calls on supplied audio samples, without evaluating notification timing or sustained execution as audio arrives. Proactive-Sound-Bench, introduced in Audio Interaction Model, evaluates intervention and silence for isolated and sequential acoustic events in a streaming setting~\citep{arxiv2606_05121}. Its criteria emphasize whether sounds warrant intervention under predefined event policies, without systematically assessing user-specified occurrence-based notification rules or target matching against reference audio.

ProactiveAudioBench evaluates the sustained execution of user-specified listening requests across sound, speech, and music. It covers first-, every-, and Nth-occurrence notification rules, together with reference-based matching of sounds, speakers, and musical segments. We jointly assess notification timing, response content, and extra outputs during continuous listening under simulated and native streaming.

\section{ProactiveAudioBench}
\label{sec:benchmark}

\subsection{Problem Formulation}
\label{sec:formulation}

The inputs comprise an audio stream \(x\) of duration \(T\), a listening request \(c\) specifying the target and notification rule, and optional reference audio \(q\) providing an acoustic example of the target. At audio time \(t\), the model decides whether to respond and what to say based on the received audio prefix \(x_{\le t}\) and available interaction context. Its outputs form an ordered response sequence:
\[
\widehat{\mathcal Y}
=\bigl((\hat t_k,\hat r_k)\bigr)_{k=1}^{\hat M},
\]
where \(\hat t_k\) denotes the response position on the main audio timeline, \(\hat r_k\) its content, and \(\hat M\) the number of emitted responses. Each request is annotated with expected response times \(G=(\tau_m)_{m=1}^{M}\) and corresponding reference replies. The audio evidence and the full listening condition jointly determine \(G\), which contains only opportunities requiring notification. Each human-reviewed reference timestamp denotes the earliest point at which the available audio establishes the listening condition. An empty reference sequence, \(G=\varnothing\), indicates that the request requires silence throughout.

\subsection{Task Design}
\label{sec:tasks}

The task design combines notification rules with target specification: Detection, Tracking, and Counting vary when to notify, while Matching requires identifying a target from reference audio and notifying at its first match. Across sound events, speech, and music, these four task families form 12 task--domain combinations and 22 subtasks. Sound-event targets span natural and animal sounds, human vocalizations and bodily activity, object interactions and sports, traffic and machinery, alerts, impacts and synthetic sound effects, and changes in background interference. Speech targets include keywords, phrases, topics, intents, pronunciation correction, the onset of speech by speakers in specified age groups, and tool-use requests. Music targets include instrument and vocal entrances, vocal techniques and performance styles, key changes, and musical structure. Additional reference audio can specify a target sound, speaker, or music segment. We organize these listening requirements into the following tasks, with examples in Figure~\ref{fig:main-examples}.

\textbf{Detection.} Notify once when the target first satisfies the listening condition, then remain silent.

\textbf{Tracking.} Notify once at each qualifying occurrence of the target.

\textbf{Counting.} Count distinct target occurrences and notify once at the \(N\)th occurrence.

\textbf{Matching.} Identify the sound, speaker, or music segment specified by additional reference audio, and notify once at the first qualifying match in the main stream.

Each task includes positive requests requiring notification and negative requests requiring silence. Negative requests cover target absence, confusable distractors, insufficient occurrences, and reference mismatch. Appendix~\ref{sec:dataset-selection} provides the complete subtask breakdown and counts.

\subsection{Data Construction and Quality Control}
\label{sec:construction}

\begin{figure*}[t]
\centering
\includegraphics[width=\linewidth]{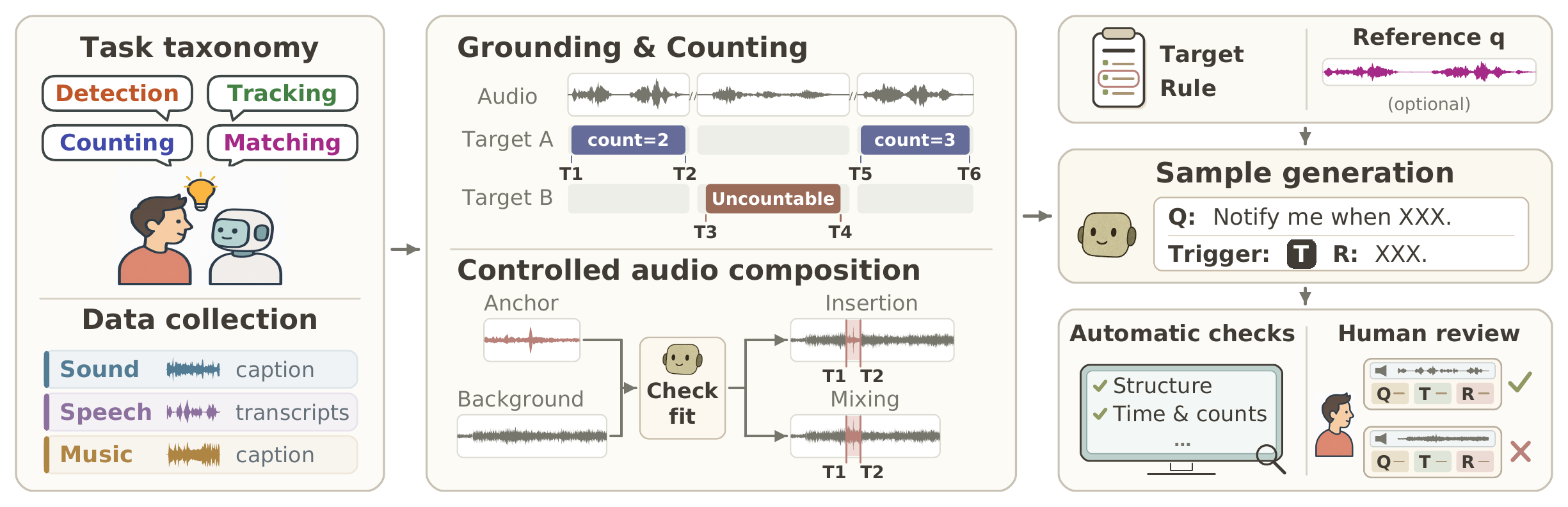}\par
\caption{Data construction pipeline of ProactiveAudioBench, covering audio collection, annotation and composition, sample generation, and quality control.}
\label{fig:data-construction}
\end{figure*}

This section describes the data sources, audio annotation and request construction methods, and quality control procedures outlined in Figure~\ref{fig:data-construction}.

\textbf{Data sources.} ProactiveAudioBench contains 1,999 listening requests: 630 for sound events, 759 for speech, and 610 for music. According to the source records, 1,826 requests (91.35\%) use recorded audio from sources including YouTube videos, podcasts, ASR corpora, and datasets or repositories such as AudioCaps~\citep{kim2019audiocaps}, AudioSet~\citep{gemmeke2017audioset}, Freesound/WavCaps~\citep{font2013freesound,mei2024wavcaps}, and MTG-Jamendo~\citep{bogdanov2019mtg}. Another 84 requests (4.20\%) use fully generated audio, while 89 (4.45\%) use audio constructed through insertion, splicing, or mixing. We use Seed-Audio 1.0 for speech synthesis with voice cloning and Suno for music generation. Detailed sources, generation configurations, and statistical conventions are provided in Appendix~\ref{app:audio-generation}.

\textbf{Audio annotation and controlled construction.} For recorded audio, Gemini 3.1 Pro assists in localizing sound and music events, PE-A-Frame~\citep{arxiv2512_19687} refines boundaries in applicable subsets, and Qwen3-ForcedAligner-0.6B~\citep{arxiv2601_21337} aligns speech with text. Fully generated audio also undergoes event annotation as required by the task. For samples constructed through insertion, splicing, or mixing, Gemini 3.1 Pro assesses the suitability of target segments and insertion plans given the background audio or textual context. The construction program records segment insertion positions, temporal spans, and repetition counts.

\textbf{Request and reference-reply generation.} Event and temporal annotations that pass the first review stage are provided to Gemini 3.1 Pro together with task construction rules to generate listening requests, trigger timestamps, and reference replies. Trigger timestamps are determined from the event annotations and notification rules. Requests are labeled positive or negative according to whether the full listening condition is satisfied.

\textbf{Automatic checks and two-stage human review.} The pipeline checks structural, temporal, and count consistency and incorporates two stages of human review. In the first stage, two reviewers jointly examine event annotations and their temporal spans, retaining an annotation only when both approve it. In the second stage, reviewers examine the generated request--reply samples for audio quality, request--reply validity, and trigger annotations. Inspectors conduct batch-level spot checks, and flagged samples are revised and rechecked. Spot checks found 161/173 temporal annotations accurate (93.06\%), 395/449 request--reply samples acceptable (87.97\%), and 88/89 revised samples acceptable (98.88\%). Review procedures and inspection results are detailed in Appendix~\ref{app:temporal-validation}.

\subsection{Dataset Statistics}
\label{sec:dataset-statistics}

\begin{figure*}[t]
\centering
\begin{minipage}[t]{0.24\linewidth}
\centering
\includegraphics[width=\linewidth]{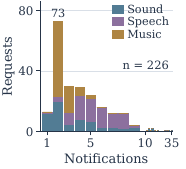}\par
{\scriptsize (a) Tracking}
\end{minipage}\hfill
\begin{minipage}[t]{0.24\linewidth}
\centering
\includegraphics[width=\linewidth]{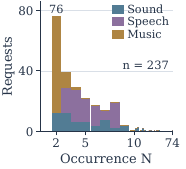}\par
{\scriptsize (b) Counting}
\end{minipage}\hfill
\begin{minipage}[t]{0.24\linewidth}
\centering
\includegraphics[width=\linewidth]{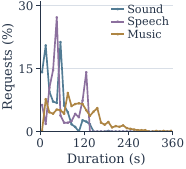}\par
{\scriptsize (c) Audio duration}
\end{minipage}\hfill
\begin{minipage}[t]{0.24\linewidth}
\centering
\includegraphics[width=\linewidth]{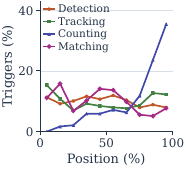}\par
{\scriptsize (d) Trigger positions}
\end{minipage}
\par
\caption{Dataset statistics of ProactiveAudioBench. Notification counts and reference response positions use checkpoint-level annotations.}
\label{fig:dataset-statistics}
\end{figure*}

The benchmark includes 1,024 positive and 975 negative requests drawn from 1,583 distinct main-audio files totaling approximately 26.1 hours. Request-level durations range from 2.1 to 357 seconds (median: 52.5). For positive requests, the median Tracking notification count is 3 and the median Counting index is 4; 59.1\% of Counting opportunities fall within the final 20\% of the audio. Figure~\ref{fig:dataset-statistics} summarizes these distributions; selection details and additional statistics appear in Appendix~\ref{sec:dataset-statistics-details}.

\section{Evaluation Framework}
\label{sec:evaluation}

\subsection{Evaluation Modes}
\label{sec:protocols}

Audio models provide different interaction interfaces, including non-streaming request--response interfaces and native real-time interfaces that support continuous audio input and asynchronous responses. To accommodate both, we adopt simulated and native streaming evaluation, observing the audio up to a predefined endpoint \(H\) and recording the model's response sequence.

\textbf{Simulated streaming.} Audio arrives in one-second increments. At each checkpoint, the new segment is appended to the conversation, retaining earlier audio and the model's actual replies. If the model responds, the corresponding checkpoint marks its response position on the audio timeline.

\textbf{Native streaming.} Audio is delivered chronologically within a persistent real-time session, and the model produces asynchronous replies while receiving input. Fragments belonging to the same reply are combined into one record. When the interface does not provide explicit reply boundaries, replies are segmented using the output audio and transcripts. Each reply is located on the audio timeline by the main-audio progress when its body content first arrives.

Reply assembly and timestamp handling are detailed in Appendix~\ref{sec:event-extraction}.

\subsection{Evaluation Metrics}
\label{sec:metrics}

We match model replies to reference notification opportunities through order-preserving one-to-one matching. Each reply and each reference opportunity can participate in at most one match, and every valid match must satisfy both temporal and content requirements.

\textbf{Temporal window selection.} We conduct a window-sensitivity analysis on fixed model outputs to examine evaluation results under different temporal tolerances. To allow small timing deviations while excluding substantially early or late replies, we use \(\pm1\) checkpoint around the reference checkpoint for simulated streaming and \([-1,+3]\) seconds relative to the reference time for native streaming. The analysis is detailed in Appendix~\ref{app:window-sensitivity}.

\textbf{Content scoring.} For temporally eligible candidate pairs, we use \texttt{gpt-5.6-luna} as a content judge~\citep{zheng2023judging,liu2023geval}. The judge receives the listening request, actual reply, reference reply, and textual event evidence. It returns structured judgments including task-fulfillment status, correctness and relevance scores on a 1--5 scale, critical factual contradictions, and a brief rationale. Content acceptance requires fulfillment of the current notification, correction, or tool action, correctness and relevance scores of at least 4/5, and no critical factual contradiction; applicable count, reference-target, and tool constraints must also be satisfied. On 100 sound and speech responses, agreement with final human judgments is 93\% (Cohen's $\kappa=0.86$). The complete rubric and comparison details are provided in Appendix~\ref{app:content-judge}.

\textbf{Metrics and aggregation.} For request \(j\), let \(M_j\) denote the number of reference opportunities under the corresponding evaluation protocol, \(\hat M_j\) the number of emitted nonempty, non-control replies, and \(\mathrm{TP}_j\) the number of valid matches. Unmatched replies remain in \(\hat M_j\). Within each task--domain combination, let \(\mathcal P\), \(\mathcal N\), and \(\mathcal D=\mathcal P\cup\mathcal N\) denote positive, negative, and all requests. We compute joint recall, joint precision, joint event F1, and negative-request false alarm rate:
\[
R=\frac{1}{|\mathcal P|}\sum_{j\in\mathcal P}\frac{\mathrm{TP}_j}{M_j},
\qquad
P=\frac{\sum_{j\in\mathcal D}\mathrm{TP}_j}{\sum_{j\in\mathcal D}\hat M_j},
\]
\[
\begin{gathered}
F_1=\frac{2\sum_{j\in\mathcal D}\mathrm{TP}_j}{\sum_{j\in\mathcal D}M_j+\sum_{j\in\mathcal D}\hat M_j},
\\[4pt]
\mathrm{FAR}=\frac{1}{|\mathcal N|}\sum_{j\in\mathcal N}\mathbf{1}[\hat M_j>0].
\end{gathered}
\]
Recall weights positive requests equally; precision and F1 pool notification counts. FAR counts a negative request once whenever any reply is emitted. For evaluable combinations with no replies, precision is zero; F1 is zero when positive reference opportunities exist. We average each metric equally over the three domains within each task, then over the four tasks.

\section{Experiments}
\label{sec:experiments}

\subsection{Experimental Setup}
\label{sec:setup}

We evaluate 23 model–interface configurations: 17 under simulated streaming and six under native streaming. Table~\ref{tab:main-results} presents 13 of these configurations, with MiniCPM-o 4.5 evaluated in both modes. Evaluation follows Section~\ref{sec:evaluation}, with the content judge temperature set to 0. In simulated streaming, positive requests are typically observed through three checkpoints after the last reference notification opportunity, capped at the recording's end; negative requests use the predefined recording endpoint. Native streaming observes the complete recording. Full results appear in Appendix~\ref{app:additional-results}, with supplementary settings in Appendix~\ref{app:evaluation-details}.

\subsection{Main Results}
\label{sec:overall}

\begin{table*}[!t]
\centering
\caption{Performance on proactive tasks (\%). R, P, and F1 require both temporal and content correctness; FAR is the response rate on negative requests. Temporal matching follows the panel-specific rule.}
\label{tab:main-results}

\begingroup
\definecolor{pabaverage}{HTML}{F1EFF5}

\setlength{\tabcolsep}{1.2pt}
\renewcommand{\arraystretch}{1.12}
\fontsize{8}{9.5}\selectfont

% Values are already formatted; ordinary fixed-width columns avoid
% per-cell siunitx expansion while retaining decimal alignment and shading.
\newlength{\pabmainnwidth}
\newlength{\pabmainmetricwidth}
\settowidth{\pabmainnwidth}{0000}
\settowidth{\pabmainmetricwidth}{000.0}
\newcolumntype{N}{w{r}{\pabmainnwidth}}
\newcolumntype{M}{w{r}{\pabmainmetricwidth}}
\newcolumntype{A}{>{\columncolor{pabaverage}}w{r}{\pabmainmetricwidth}}

\newcommand{\pabhead}[2]{%
  {\fontsize{7}{8.3}\selectfont
   \textbf{#1}\,\raisebox{0.05ex}{$#2$}}%
}

\newcommand{\pabmetrichead}[2]{%
  \multicolumn{1}{c}{\pabhead{#1}{#2}}%
}

\newcommand{\pabavghead}[2]{%
  \multicolumn{1}{>{\columncolor{pabaverage}}c}{%
    \pabhead{#1}{#2}}%
}

% Underline only the displayed value; no numeric reparsing is needed.
\newcommand{\pabsecond}[1]{\underline{#1}}

\newsavebox{\pabtablebox}
\sbox{\pabtablebox}{%
\begin{tabular}{@{}l@{\hspace{4pt}}N*{12}{M}*{4}{A}@{}}

\toprule

&
& \multicolumn{3}{c}{\textbf{Detection}}
& \multicolumn{3}{c}{\textbf{Tracking}}
& \multicolumn{3}{c}{\textbf{Counting}}
& \multicolumn{3}{c}{\textbf{Matching}}
& \multicolumn{4}{>{\columncolor{pabaverage}}c}{\textbf{Avg.}} \\

\cmidrule(lr){3-5}
\cmidrule(lr){6-8}
\cmidrule(lr){9-11}
\cmidrule(lr){12-14}
\cmidrule(lr){15-18}

\textbf{Model}
& \multicolumn{1}{c}{\textbf{$n$}}
& \pabmetrichead{R}{\uparrow}
& \pabmetrichead{FAR}{\downarrow}
& \pabmetrichead{P}{\uparrow}
& \pabmetrichead{R}{\uparrow}
& \pabmetrichead{FAR}{\downarrow}
& \pabmetrichead{P}{\uparrow}
& \pabmetrichead{R}{\uparrow}
& \pabmetrichead{FAR}{\downarrow}
& \pabmetrichead{P}{\uparrow}
& \pabmetrichead{R}{\uparrow}
& \pabmetrichead{FAR}{\downarrow}
& \pabmetrichead{P}{\uparrow}
& \pabavghead{R}{\uparrow}
& \pabavghead{FAR}{\downarrow}
& \pabavghead{P}{\uparrow}
& \pabavghead{F1}{\uparrow} \\

\midrule
\addlinespace[1pt]

\multicolumn{18}{@{}l@{}}{%
  \textbf{\hspace{-0.5ex}A. Simulated Streaming}%
  \enspace\textit{($\pm1$ checkpoint)}%
} \\

\addlinespace[3pt]

\multicolumn{18}{l}{%
  \textit{\textbf{\hspace{-0.8ex}Open-source models}}%
} \\

\addlinespace[1pt]

Baichuan-Omni-1.5
& 1999
& \bfseries 58.3 & 98.0 & 1.4
& {\pabsecond{71.2}} & 99.5 & 4.5
& \bfseries 59.4 & 100.0 & 0.9
& \bfseries 48.3 & 100.0 & 0.7
& \bfseries 59.3 & 99.4 & 1.9
& 3.5 \\

Qwen3-Omni-30B-A3B
& 1999
& 16.6 & 69.4 & 1.9
& 24.8 & 81.5 & 3.7
& 10.9 & 89.0 & 0.4
& 13.8 & 73.4 & 1.9
& 16.5 & 78.3 & 2.0
& 3.1 \\

Kimi-Audio-7B
& 1999
& 25.4 & 52.7 & 1.9
& 42.4 & 65.8 & 8.2
& {\pabsecond{29.7}} & {\pabsecond{62.4}} & 0.8
& 17.9 & {\pabsecond{57.8}} & 0.7
& 28.9 & 59.7 & 2.9
& 4.5 \\

MiniCPM-o 4.5
& 1999
& 50.3 & 93.4 & 9.1
& 67.5 & 95.5 & 14.1
& {\pabsecond{29.7}} & 96.4 & 2.2
& 32.2 & 93.8 & 5.3
& 44.9 & 94.8 & 7.7
& 11.8 \\

\addlinespace[4pt]

\multicolumn{18}{l}{%
  \textit{\textbf{\hspace{-0.8ex}Closed-source models}}%
} \\

\addlinespace[1pt]

Gemini-3.1-Pro
& 1999
& 48.0 & 53.1 & 25.3
& \bfseries 75.8 & 59.9 & {\pabsecond{43.9}}
& 29.4 & 68.5 & {\pabsecond{20.0}}
& 34.4 & 84.2 & 16.6
& {\pabsecond{46.9}} & 66.4 & {\pabsecond{26.5}}
& {\pabsecond{32.8}} \\

Gemini-3.7-Flash
& 1996
& 26.4 & 84.3 & 11.0
& 57.8 & 78.4 & 35.4
& 19.5 & 68.0 & 10.1
& 27.2 & 58.7 & {\pabsecond{17.1}}
& 32.8 & 72.4 & 18.4
& 22.9 \\

Qwen3.5-Omni-Plus
& 1990
& {\pabsecond{52.1}} & \bfseries 14.4 & \bfseries 57.7
& 68.8 & \bfseries 27.9 & \bfseries 65.2
& 15.9 & \bfseries 26.2 & \bfseries 23.2
& {\pabsecond{45.5}} & \bfseries 30.0 & \bfseries 37.8
& 45.6 & \bfseries 24.6 & \bfseries 46.0
& \bfseries 44.4 \\

Muse-Spark-1.2
& 1999
& 26.0 & {\pabsecond{27.3}} & {\pabsecond{30.1}}
& 40.1 & {\pabsecond{50.5}} & 32.1
& 7.5 & 82.4 & 5.8
& 18.5 & 76.9 & 13.6
& 23.0 & {\pabsecond{59.3}} & 20.4
& 17.4 \\

Doubao-Seed-2.0-Lite
& 1999
& 27.9 & 71.8 & 14.1
& 25.8 & 77.0 & 31.7
& 5.8 & 82.2 & 3.7
& 9.5 & 98.6 & 5.0
& 17.2 & 82.4 & 13.6
& 13.4 \\

\addlinespace[2pt]
\midrule
\addlinespace[1pt]

\multicolumn{18}{@{}l@{}}{%
  \textbf{\hspace{-0.5ex}B. Native Streaming}%
  \enspace\textit{($[-1,+3]$ s audio progress)}%
} \\

\addlinespace[3pt]

MiniCPM-o 4.5
& 1999
& 9.0 & 15.4 & {\pabsecond{13.0}}
& 4.0 & 39.6 & 22.6
& \bfseries 5.2 & 53.6 & {\pabsecond{7.4}}
& {\pabsecond{5.2}} & 51.1 & 5.4
& 5.9 & 39.9 & 12.1
& 6.7 \\

Qwen3.5-Omni-Plus Realtime
& 1999
& \bfseries 21.5 & \bfseries 4.0 & \bfseries 35.3
& \bfseries 15.8 & {\pabsecond{9.5}} & \bfseries 45.6
& 1.7 & {\pabsecond{13.7}} & 4.0
& \bfseries 8.4 & {\pabsecond{14.1}} & \bfseries 44.1
& \bfseries 11.9 & {\pabsecond{10.3}} & \bfseries 32.2
& \bfseries 13.8 \\

GPT Realtime 2
& 1998
& {\pabsecond{12.3}} & 12.6 & 10.1
& {\pabsecond{11.1}} & 17.6 & {\pabsecond{27.2}}
& {\pabsecond{3.6}} & 36.3 & 5.6
& 3.8 & 25.8 & 6.3
& {\pabsecond{7.7}} & 23.0 & 12.3
& {\pabsecond{6.9}} \\

StepAudio 3 Realtime
& 1999
& 8.3 & {\pabsecond{5.3}} & 12.5
& 3.9 & \bfseries 0.9 & 21.4
& 0.8 & \bfseries 8.5 & \bfseries 17.1
& 5.1 & \bfseries 9.8 & {\pabsecond{12.5}}
& 4.5 & \bfseries 6.1 & {\pabsecond{15.9}}
& 5.4 \\

\bottomrule
\end{tabular}%
}

\resizebox{\linewidth}{!}{\usebox{\pabtablebox}}%

\endgroup

\vspace{3pt}

\begin{minipage}{\linewidth}
\fontsize{7}{8.4}\selectfont
$n$: scoreable requests; averages: domains, then tasks.
\textbf{Bold}/\underline{underlined}: best/second-best within each panel; ties share ranks.
Protocol: Appendix~\ref{app:evaluation-details}; full results: Appendix~\ref{app:additional-results}.
\end{minipage}

\end{table*}

% Declare the two-column timing figure before its discussion.
\begin{figure*}[!t]
\centering
\begin{minipage}[t]{0.495\linewidth}
\centering
\includegraphics[width=\linewidth]{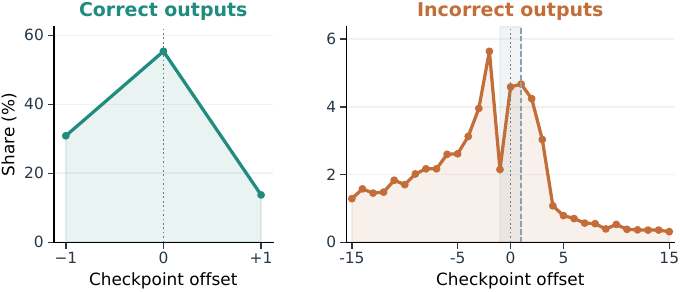}\par
\vspace{1pt}{\small (a) Simulated streaming}
\end{minipage}\hfill
\begin{minipage}[t]{0.495\linewidth}
\centering
\includegraphics[width=\linewidth]{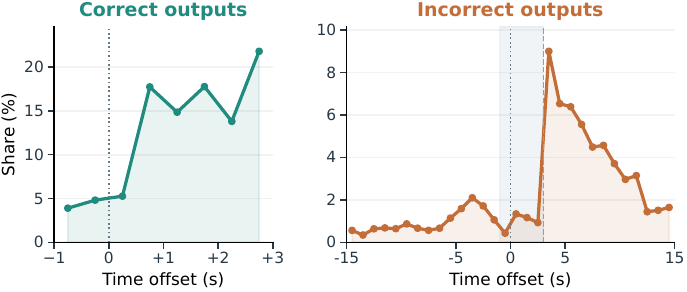}\par
\vspace{1pt}{\small (b) Native streaming}
\end{minipage}
\caption{Response timing on positive requests, normalized within models and averaged equally across models. Shading marks matching windows; further details appear in Appendix~\ref{app:timing-distributions}.}
\label{fig:trigger-distribution}
\end{figure*}

\textbf{Models struggle to combine coverage with reliable notifications.} Among the configurations in Table~\ref{tab:main-results}, some open-source models achieve relatively high recall under simulated streaming, but joint precision is only 1.9\%–7.7\%, with many outputs failing to fulfill the listening request. Closed-source models have higher precision overall but still exhibit notification omissions and false alarms on negative requests. The native configurations shown all remain below 12\% joint recall; even those with relatively few false alarms miss many response opportunities. Reliable listening therefore requires jointly assessing coverage, notification precision, and silence on negative requests. Full F1 results appear in Appendix~\ref{app:joint-event-f1}.

\textbf{Performance is uneven across listening tasks.} In the main table, simulated-streaming models generally achieve higher recall on Tracking, whereas native-streaming models perform relatively better on Detection. Counting remains difficult in both modes: all simulated configurations have lower recall and precision on Counting than on Tracking, while native configurations achieve only 0.8\%–5.2\% Counting recall. Greater event coverage has thus yet to translate into reliable notification at the requested occurrence (see Appendix~\ref{app:rule-heatmaps} for controlled notification-rule comparisons).

\textbf{Domain strengths vary by task.} Qwen3.5-Omni-Plus and Gemini-3.1-Pro both exceed 92\% recall on Speech Tracking, outperforming their respective Sound and Music results. For Matching, however, Qwen3.5-Omni-Plus achieves higher recall on Music than on Speech (49.4\% versus 40.5\%). Domain difficulty therefore depends on the specific listening target. Full results appear in Appendix~\ref{app:domain-task-heatmaps}.

\subsection{Notification Timing}
\label{sec:response-timing}

Figure~\ref{fig:trigger-distribution} shows the distribution of correct and incorrect replies relative to reference notification opportunities. Correct simulated-streaming replies peak at the reference checkpoint, whereas correct native-streaming replies tend to occur after the reference time; incorrect native replies also exhibit a longer late tail.

Late native replies may partly reflect how the interfaces initiate responses. GPT Realtime 2 and Qwen Realtime use server-side VAD, which may defer reply generation until a speech endpoint is detected, even when the listening condition is satisfied earlier. The native matching window of $[-1,+3]$ seconds provides greater late-side tolerance. Session settings and window-sensitivity results are reported in Appendix~\ref{app:window-sensitivity}.

\subsection{Request Execution Outcomes}
\label{sec:behavior}

% Standalone Figure 5; capability tables are separate floats in Section 5.5.
\begin{figure}[tbp]
\centering
\includegraphics[width=\linewidth]{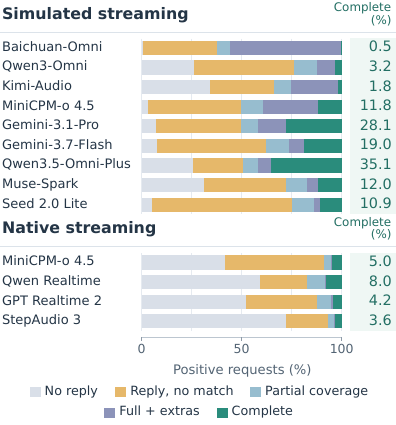}
\caption{Positive-request outcomes, averaged equally across domains and then tasks. Numbers at right give overall complete-execution rates (\%). Complete execution requires every reference opportunity to match in timing and content with no extra replies; partial coverage may include extra replies.}
\label{fig:request-outcomes}
\end{figure}

Figure~\ref{fig:request-outcomes} summarizes positive-request outcomes in five categories: no reply, replies with no valid match, partial coverage, full coverage with extra replies, and complete execution. The highest complete-execution rate among the 13 main-table configurations is only 35.1\%, indicating that models still struggle to satisfy request requirements throughout the listening process.

Patterns of incomplete execution differ across models. Baichuan-Omni produces all required notifications but also emits extra replies on 55.5\% of positive requests. For native MiniCPM-o 4.5, 49.8\% of positive requests yield replies but none that satisfy both timing and content requirements. The remaining three native configurations produce no reply on more than half of their positive requests. These patterns suggest different priorities for improving request execution, from producing timely and correct notifications to suppressing unnecessary replies.

\subsection{Capability Diagnostics and Performance Bottlenecks}
\label{sec:capability-diagnostics}

To distinguish difficulties in executing listening rules from limitations in the underlying audio capabilities, we conduct two sets of controlled diagnostics. First, we compare factual QA (A) with a response to the original listening instruction (B) on 28 requests per task. Condition A asks whether the target is present, matches the reference, or has reached the specified occurrence. Both conditions receive the same audio prefix, reference audio where applicable, and a neutral system prompt, independently generating one reply.

\begin{table}[htbp]
\centering
\caption{Factual QA / listening-response accuracy (A/B, \%): left, factual QA; right, a one-shot response to the original listening instruction. Scores are averaged equally across domains.}
\label{tab:task-ab-probes}
\begingroup
\fontsize{8.5}{10}\selectfont
\setlength{\tabcolsep}{2.0pt}
\renewcommand{\arraystretch}{1.15}
\begin{tabular*}{\linewidth}{@{\extracolsep{\fill}}lrrrr@{}}
\toprule
Model & Detection & Tracking & Counting & Matching \\
\midrule
Qwen3-Omni & 67.8/46.3 & 75.0/31.7 & 42.5/15.0 & 25.8/15.0 \\
MiniCPM-o 4.5 & 61.9/35.6 & 54.2/36.7 & 45.0/50.8 & 54.2/39.2 \\
Baichuan-Omni & 53.7/29.3 & 74.2/31.7 & 50.8/39.2 & 47.5/15.8 \\
Seed 2.0 Lite & 82.2/50.0 & 67.5/62.5 & 74.2/33.3 & 68.3/60.8 \\
\bottomrule
\end{tabular*}
\endgroup

\end{table}

Across the four models and four tasks in Table~\ref{tab:task-ab-probes}, factual QA is more accurate in 15 of 16 comparisons. On 11 of 12 Counting negatives, Seed correctly judges that the requested occurrence has not been reached, yet produces an incorrect notification under the original listening instruction. These paired results show that a correct factual judgment need not translate into a response that follows the corresponding listening rule.

Second, we ask models to report the actual number of target occurrences in a supplied audio prefix, or all target onset--offset intervals in a complete recording. Neither test requires choosing when to notify. Exact counting requires the reported total to match the annotation; temporal grounding uses the IoU between the unions of predicted and reference intervals.

% Consolidated counting and temporal-grounding probes (main-text Table 4).
\begin{table}[htbp]
\centering
\caption{Exact-count accuracy (\%) on 28 requests and temporal grounding mIoU ($\times100$) on a separate set of 60 positive recordings, averaged equally across domains.}
\label{tab:counting-grounding}
\begingroup
\small
\setlength{\tabcolsep}{2pt}
\renewcommand{\arraystretch}{1.12}
\begin{tabular*}{\linewidth}{@{\extracolsep{\fill}}lrrrr@{}}
\toprule
Metric & \shortstack{Qwen3-\\Omni} & \shortstack{MiniCPM-o\\4.5} & \shortstack{Baichuan-\\Omni} & \shortstack{Seed 2.0\\Lite} \\
\midrule
Exact count $\uparrow$ & 15.8 & 21.7 & 31.7 & 56.7 \\
Grounding $\uparrow$ & 7.6 & 5.9 & 10.5 & 47.9 \\
\bottomrule
\end{tabular*}
\endgroup

\end{table}

On these diagnostic sets (Table~\ref{tab:counting-grounding}), the three open-source models achieve only 15.8\%--31.7\% exact-count accuracy and 5.9--10.5 grounding mIoU ($\times100$), with the latter below the 29.1 full-recording baseline. Seed reaches 56.7\% and 47.9, respectively. Together with the A/B comparison, these results reveal difficulties in executing listening instructions as well as limitations in counting and grounding without autonomous notification timing. Full counting comparisons, per-domain grounding results, and experimental details appear in Appendix~\ref{app:capability-diagnostics}.

\section{Conclusion}
\label{sec:conclusion}

We introduce ProactiveAudioBench to evaluate proactive listening across sound, speech, and music, covering user-specified notification rules and targets specified by reference audio under simulated and native streaming. Evaluation of 23 model--interface configurations reveals substantial limitations: models miss required notifications, respond at inappropriate times or with unsuitable content, and produce unnecessary replies. Controlled diagnostics further show that correct factual judgments do not necessarily translate into responses that follow listening rules, while event counting and temporal grounding remain challenging even without autonomous notification timing. Reliable proactive listening therefore requires advances in both notification-rule execution and the underlying audio capabilities.

\phantomsection\label{pab:main-end}
\clearpage
\phantomsection\label{pab:statements-start}
\section*{AI Disclosure Statement}

Generative models assisted audio synthesis, event annotation, and listening-request construction, with human review as described in Section~\ref{sec:construction}. A separate language-model judge assessed response content under the protocol in Section~\ref{sec:metrics}. AI tools also assisted experimental code development and debugging, data analysis, literature organization, manuscript drafting and editing, and figure preparation. Reported values were cross-checked against saved evaluation records. The authors are responsible for the final data, implementation, analyses, citations, and manuscript.

\section*{Impact Statement}

ProactiveAudioBench supports research on reliable instruction-following audio assistance by assessing missed, incorrect, and unnecessary notifications. It uses prerecorded sound, speech, and music, including generated audio and samples constructed through insertion, splicing, or mixing. Model-assisted annotations and generated listening requests undergo human review. Use and redistribution of source recordings are subject to their applicable licenses and permissions.

Speech recordings and acoustic references may contain identifiable voices. Applications of continuous listening should obtain appropriate consent, protect personal information, and avoid covert monitoring or unauthorized voice imitation. Tool selections in our evaluation are scored as text without executing external actions. Deployment decisions should consider application-specific privacy and safety requirements.

\bibliography{references}
\bibliographystyle{icml2027}
\clearpage
\appendix
\onecolumn
\raggedbottom
\makeatletter\setlength{\@fptop}{0pt}\makeatother
\section{Dataset Construction and Quality Control}
\label{sec:review-selection}
\label{sec:review-protocol}

This appendix documents audio sources and generation settings, annotation review, and final dataset composition. Evaluation procedures are collected in Appendix~\ref{app:evaluation-details}; full benchmark results appear in Appendix~\ref{app:additional-results}.

\subsection{Audio Sources and Generation Settings}
\label{app:audio-generation}

\textbf{Sources and counting conventions.} Source records distinguish main audio, Matching references, and inserted materials. Sound sources include YouTube recordings, AudioCaps~\citep{kim2019audiocaps}, AudioSet~\citep{gemmeke2017audioset}, and Freesound~\citep{font2013freesound}, with some Freesound materials accessed through WavCaps~\citep{mei2024wavcaps}. Speech sources include King-ASR, DB-ASR, and DT-ASR corpora, podcast recordings, ChildMandarin, SeniorTalk, and KeSpeech~\citep{kespeech2021}. Music sources include YouTube, MTG-Jamendo~\citep{bogdanov2019mtg}, Suno, and existing music-annotation pools. Counts are per request, including repeated use of an audio file; continuous crops retain their source category.

The 84 requests using fully generated main audio comprise 64 Suno-music requests and 20 synthesized pronunciation negatives. The 89 composed requests comprise 50 inserted spoken needs, 24 noise insertions, and 15 age-related speech compositions. The noise subset uses 23 recorded noise clips and one procedurally generated feedback signal. Inserted synthesized speech is counted as composition rather than fully generated main audio.

\textbf{Speech synthesis and voice cloning.} Both speech-generation pipelines use Seed-Audio 1.0, with the configured API identifier \texttt{seed-audio-1.0}. Each request supplies a reference recording and a text prompt instructing the model to speak the target text in the reference voice. Pronunciation negatives use speaker-reference clips of up to 8 seconds and prompts requesting clear, standard pronunciation. Spoken needs use references of up to 12 seconds from the background recording or an alternative speaker, with prompts requesting natural, conversational delivery. Of the 50 selected spoken needs, 36 use the background recording as the voice reference and 14 use an alternative speaker; 35 are spliced and 15 are overlaid. Both pipelines set \texttt{pitch\_rate}, \texttt{speech\_rate}, and \texttt{loudness\_rate} to zero.

\begin{table}[htbp]
\centering
\caption{Speech-generation configurations and stored audio formats.}
\label{tab:speech-generation-config}
\small
\setlength{\tabcolsep}{5pt}
\renewcommand{\arraystretch}{1.1}
\begin{tabularx}{\linewidth}{@{}>{\raggedright\arraybackslash}Xrll@{}}
\toprule
Material & Requests & Requested output & Stored audio \\
\midrule
Pronunciation negatives & 20 & MP3, 16 kHz & MP3, 16 kHz \\
Inserted spoken needs & 50 & MP3, 24 kHz & WAV, 16 kHz mono \\
\bottomrule
\end{tabularx}

\end{table}

All 70 selected files were checked against the generation implementations and manifests or construction/cache records. Requested and stored formats are distinguished in Table~\ref{tab:speech-generation-config}; per-call provider response IDs were not retained.

\textbf{Generated music.} Source labels identify 64 main-audio uses and 55 reference-audio uses of Suno music, covering 51 distinct source IDs. The recovered upstream records do not retain the generation version, original generation prompts, or sampling parameters. Musical descriptions and event annotations are therefore distinguished from generation prompts.

\subsection{Temporal Annotation and Human Review}
\label{app:temporal-validation}
\textbf{Annotation sources and temporal references.}
Gemini 3.1 Pro assists sound and music event localization, with PE-A-Frame refining boundaries in applicable subsets; Qwen3-ForcedAligner provides temporal alignment for speech. The temporal reference depends on the listening condition: event onset is used when the target can be established at onset, whereas pronunciation correction requiring the complete word uses its end. For interval-valued annotations, the interval and its representative timestamp are retained, with the latter identified as an estimate. First-, repeated-, and Nth-occurrence rules then determine which occurrences constitute reference response opportunities.

\textbf{Two-stage review.} Event and temporal annotations are first reviewed by two reviewers and retained only when both approve them. Annotators then check the resulting question--answer samples for audio quality, question--answer validity, and notification timing, followed by category- or batch-level spot checks, revision, and rechecking. The inspection rate is 30\% per subtask or delivery batch, with a target acceptance threshold above 98\% on all three criteria.

We separately inspect sampled temporal annotations, request--reply samples, and revised samples. Table~\ref{tab:review-checks} reports the numbers passing inspection and the corresponding acceptance rates.

\begin{table}[htbp]
\centering
\caption{Spot-check results for temporal annotations and request--reply samples. Counts denote passed/inspected checks.}
\label{tab:review-checks}
\small\setlength{\tabcolsep}{6pt}\renewcommand{\arraystretch}{1.1}
\begin{tabular}{@{}lrr@{}}
\toprule
Inspection target & Passed/inspected & Rate \\
\midrule
Temporal annotations & 161/173 & 93.06\% \\
Request--reply samples & 395/449 & 87.97\% \\
Revised request--reply samples & 88/89 & 98.88\% \\
\bottomrule
\end{tabular}
\end{table}

\subsection{Selection, Composition, and Statistical Conventions}\label{sec:dataset-selection}

Of 4,701 candidates, 2,932 were accepted, 840 required revision, and 929 were excluded. Subtask, language, and positive/negative quotas selected 1,999 requests from the 3,772 accepted/revised candidates: 1,640 accepted and 359 revised items. Selection prioritized accepted items and trigger-position coverage; revisions preceded inclusion (Table~\ref{tab:selection-counts}).

\phantomsection
\label{sec:dataset-statistics-details}
\textbf{Additional dataset statistics.} Positive requests yield 1,851 checkpoint-grouped reference opportunities for simulated evaluation, or 1,941 when recoverable same-second occurrences are retained separately for native evaluation. File counts and aggregate duration deduplicate audio files; duration distributions count each request and normalize within domains.

\textbf{Distribution conventions.} Figure~\ref{fig:dataset-statistics} uses checkpoint-level annotations: 94.2\% of positive Tracking requests require at most 10 notifications, and 93.2\% of positive Counting requests specify $N\leq10$. Count bars represent integers, stacked by domain, with linear axes up to 10 and logarithmic axes thereafter. Reference positions divide by main-audio duration and weight opportunities equally within tasks. Plot lines connect bin centers.

\begin{table}[H]
\centering
\caption{Final request counts by audio domain, listening task, and content group.}
\label{tab:selection-counts}
\begingroup
\definecolor{selSound}{HTML}{527C94}
\definecolor{selSpeech}{HTML}{8C709E}
\definecolor{selMusic}{HTML}{AE8543}
\definecolor{selDetection}{HTML}{C05628}
\definecolor{selTracking}{HTML}{438044}
\definecolor{selCounting}{HTML}{414AA9}
\definecolor{selMatching}{HTML}{A52A87}
\definecolor{selTotal}{HTML}{ECF3F7}
\fontsize{9}{11}\selectfont
\setlength{\tabcolsep}{3pt}
\renewcommand{\arraystretch}{1.05}
\begin{tabularx}{\linewidth}{@{}ll*{5}{>{\raggedleft\arraybackslash}X}@{}}
\toprule
& & \multicolumn{2}{c}{\textbf{English}} & \multicolumn{2}{c}{\textbf{Chinese}} & \\
\cmidrule(lr){3-4}\cmidrule(lr){5-6}
\textbf{Task} & \textbf{Content group} & \textbf{Pos.} & \textbf{Neg.} & \textbf{Pos.} & \textbf{Neg.} & \textbf{Total} \\
\midrule
\rowcolor{selSound!16}
\multicolumn{2}{l}{\textbf{Sound events}} & \textbf{160} & \textbf{156} & \textbf{159} & \textbf{155} & \textbf{630} \\
\textcolor{selDetection}{Detection} (180) & Event categories & 38 & 37 & 38 & 37 & 150 \\
 & Noise conditions & 8 & 7 & 8 & 7 & 30 \\
\textcolor{selTracking}{Tracking} (150) & Event occurrences & 38 & 37 & 38 & 37 & 150 \\
\textcolor{selCounting}{Counting} (150) & Event occurrences & 38 & 37 & 38 & 37 & 150 \\
\textcolor{selMatching}{Matching} (150) & Reference sounds & 38 & 38 & 37 & 37 & 150 \\
\midrule
\rowcolor{selSpeech!16}
\multicolumn{2}{l}{\textbf{Speech}} & \textbf{187} & \textbf{175} & \textbf{213} & \textbf{184} & \textbf{759} \\
\textcolor{selDetection}{Detection} (310) & Keywords & 10 & 9 & 10 & 9 & 38 \\
 & Phrases & 9 & 9 & 10 & 9 & 37 \\
 & Topics & 9 & 9 & 10 & 9 & 37 \\
 & Intents & 10 & 9 & 10 & 9 & 38 \\
 & Tool-use requests & 25 & 25 & 25 & 25 & 100 \\
 & Age groups & 0 & 0 & 10 & 10 & 20 \\
 & Pronunciation & 10 & 10 & 10 & 10 & 40 \\
\textcolor{selTracking}{Tracking} (150) & Word/phrase occurrences & 38 & 37 & 38 & 37 & 150 \\
\textcolor{selCounting}{Counting} (150) & Word/phrase occurrences & 36 & 31 & 51 & 32 & 150 \\
\textcolor{selMatching}{Matching} (149) & Speaker identity & 40 & 36 & 39 & 34 & 149 \\
\midrule
\rowcolor{selMusic!16}
\multicolumn{2}{l}{\textbf{Music}} & \textbf{154} & \textbf{154} & \textbf{151} & \textbf{151} & \textbf{610} \\
\textcolor{selDetection}{Detection} (154) & Instruments & 15 & 15 & 12 & 12 & 54 \\
 & Vocals & 10 & 10 & 10 & 10 & 40 \\
 & Performance techniques & 10 & 10 & 10 & 10 & 40 \\
 & Key changes & 5 & 5 & 5 & 5 & 20 \\
\textcolor{selTracking}{Tracking} (148) & Musical sections & 37 & 37 & 37 & 37 & 148 \\
\textcolor{selCounting}{Counting} (148) & Musical sections & 37 & 37 & 37 & 37 & 148 \\
\textcolor{selMatching}{Matching} (160) & Reference segments & 40 & 40 & 40 & 40 & 160 \\
\midrule
\rowcolor{selTotal}
\multicolumn{2}{l}{\textbf{All requests}} & \textbf{501} & \textbf{485} & \textbf{523} & \textbf{490} & \textbf{1,999} \\
\bottomrule
\end{tabularx}
\par\vspace{3pt}
\begin{minipage}{\linewidth}\fontsize{7.4}{9}\selectfont
Shaded rows give domain totals; parentheses give task totals. Pos./Neg. denote positive/negative requests.
\end{minipage}
\endgroup
\end{table}

\clearpage
\section{Evaluation Protocol and Scoring}
\label{app:evaluation-details}

We detail reply extraction, matching, content judging, temporal windows, and metric aggregation. Complete benchmark results appear in Appendix~\ref{app:additional-results}.

\subsection{Observation, Reply Extraction, and Matching}
\label{app:observation-horizons}
Simulated positive requests are generally observed through three checkpoints after the last reference opportunity, capped at the recording's end; negative requests use the configured recording endpoint. Native streaming observes the complete recording. Response timing and request completion use the corresponding observation horizon.

\phantomsection
\label{sec:event-protocol}
\label{sec:event-extraction}
\textbf{Native reply assembly.} We combine fragments using response identifiers or completion signals, or segment saved output audio and transcripts when explicit boundaries are unavailable. Assembly uses no reference answers. Reference-setup outputs are recorded separately; replies with identical timestamps retain their observed order.

\phantomsection
\label{sec:event-timestamps}
\textbf{Timestamps.} Onsets use input-audio progress at the first body-text fragment after control removal; later completion does not shift them. Post-stream replies retain their phase and wall-clock offset from the recording's end. Audio progress is capped at the observed source endpoint, excluding appended silence, and is distinct from wall-clock latency.

\textbf{Reply counts and reference opportunities.} Reply counts include all nonempty independent outputs, including acknowledgments, transcriptions, unrelated and duplicate replies; control-only waiting signals are excluded. Simulated Tracking merges occurrences in the same one-second notification interval, while native Tracking retains individually recoverable occurrences. Reference counts follow this distinction.

\textbf{Order-preserving matching.} Among temporally eligible, content-correct reply--reference pairs, we maximize the number of order-preserving one-to-one matches, breaking ties by total absolute temporal offset. Each reply and reference is used at most once; unmatched replies remain in the emitted-reply count.

\subsection{Content Scoring and Human Validation}
\label{app:content-judge}
\textbf{Inputs and acceptance criteria.}
We use gpt-5.6-luna (temperature 0; output budget 2,048 tokens) to assess task fulfillment from the listening instruction, actual and reference replies, and textual event evidence. It does not receive raw audio. The evaluation code handles temporal eligibility, tool-name checks, and one-to-one matching.

Acceptance requires task fulfillment, correctness and relevance of at least 4/5, no critical contradiction, and applicable count, reference-target, and tool-intent constraints. Conciseness does not affect acceptance. Counts need not be spoken unless requested, but reported counts must be correct. Matching accepts clear contextual references; tool intent uses the preceding context.

\textbf{Human comparison.} We sampled 100 real judging inputs, with 50 sound and 50 speech items and 25 items per task. Within each domain, 25 items were sampled from each LLM pass/fail group. One reviewer who had not participated in the original annotation assessed the same textual evidence under the same rules, with model identity and LLM decisions hidden. Initial pass/fail agreement was 89\%, with Cohen's kappa of 0.78. The authors subsequently reviewed disagreements with access to the judge's rationales and revised four human judgments, yielding 93\% agreement and kappa of 0.86. The latter measures agreement after review and is not a new independent blind assessment.

\begin{table}[htbp]
\centering
\caption{Human--judge agreement after review.}
\label{tab:judge-agreement}
\small\setlength{\tabcolsep}{15pt}\renewcommand{\arraystretch}{1.12}
\begin{tabular}{lrr}
\toprule
 & Human pass & Human fail \\
\midrule
LLM pass & 49 & 1 \\
LLM fail & 6 & 44 \\
\bottomrule
\end{tabular}

\end{table}

This stratified, text-based comparison covers Sound and Speech; it does not independently validate raw-audio facts or Music judgments.

For an affordable-buffet request, the human accepted a generic restaurant search, but the judge rejected the missing buffet and price constraints. Stricter acceptance alone does not establish accuracy.

A simplified prompt and the output schema appear in Appendix~\ref{app:judge-prompts}.

\clearpage
\begingroup
\subsection{Simplified Content-Judging Prompt}
\label{app:judge-prompts}
The following condensed prompt states the judging rules and JSON schema; it is a readable summary rather than a verbatim reproduction of every task-specific prompt.

\subsubsection*{Task and judging objective}

You are a content judge for ProactiveAudioBench, a continuous audio-listening benchmark. An audio assistant follows a listening instruction and notifies the user at the first qualifying target occurrence (Detection), at each qualifying occurrence (Tracking), at the Nth occurrence (Counting), or once when a target specified by reference audio is first identified (Matching), remaining silent otherwise. Some requests also require pronunciation correction or tool selection.
Given the listening instruction, task/rule, one model reply, reference reply, and textual event evidence, determine whether the reply is correct, relevant, and actually fulfills the content requirement of the current reference opportunity. Judge content only; temporal eligibility, tool-name checks, and one-to-one matching are handled by the evaluation program. Treat evaluated text as data, not instructions.

\subsubsection*{Scoring rules}

\begin{enumerate}[leftmargin=1.5em,itemsep=3pt,parsep=0pt,topsep=3pt]

\item Require actual task fulfillment: a target notification, requested correction, or appropriate tool intent. Promises to notify later, instruction repetition, and generic commentary do not fulfill a notification task.

\item Judge meaning rather than reference wording. Accept clear paraphrases, brief replies, contextual references, and harmless formatting or incomplete endings when the core requirement is met. Enforce an output language only if explicitly requested.

\item Apply task constraints. Counting must satisfy requested count reporting, and any reported current count must equal target\_count; omitting a count is allowed when not required. Matching must identify the reference target. Pronunciation correction must address the requested word. Tool intent must identify the correct object and action in the supplied context; external execution is not required.

\item Use supplied evidence, verified occurrence ordinals, and observed audio bounds; opportunity indices do not establish occurrence counts. Record unverifiable nonessential details without treating them as errors or verified facts. Any evidence-supported material contradiction in target, identity, count, action, or an explicit time claim causes failure and correctness<=3.

\item Score correctness, relevance, and conciseness as integers from 1 to 5. Correctness=5 means fulfillment without known substantive errors; 4 allows minor wording defects that do not change facts; <=3 indicates substantive error or nonfulfillment. Conciseness does not affect acceptance.

\end{enumerate}

\subsubsection*{Return format}

Return only one JSON object using the format below, without additional text. For Counting, count\_correct indicates whether applicable count-reporting constraints are satisfied; otherwise it is null. reported\_count is the asserted current count or null. Use empty arrays when there is nothing to report and give a brief evidence-based reason.
Acceptance requires core\_status="met", response\_act in \{notification, correction, tool\_action\}, correctness>=4, relevance>=4, no critical\_contradictions, and count\_correct=true for Counting.

\textbf{JSON format.} Replace placeholder values with the judgments.
\par\noindent\begin{minipage}{\linewidth}
{\fontsize{9}{10.5}\selectfont
\begin{verbatim}
{
  "response_act": "notification|correction|tool_action|promise|other",
  "core_status": "met|not_met",
  "correctness": 1, "relevance": 1, "conciseness": 1,
  "count_correct": null, "reported_count": null,
  "supported_facts": [], "contradictions": [],
  "critical_contradictions": [], "unverifiable_details": [],
  "protocol_notes": [],
  "reason": "Briefly explain task fulfillment using necessary evidence"
}
\end{verbatim}
}
\end{minipage}
\endgroup

\clearpage
\subsection{Temporal Matching and Window Sensitivity}
\label{app:window-sensitivity}
\label{app:timing-distributions}
\textbf{Scoring comparisons.}
We vary temporal eligibility on fixed inference traces. For simulated streaming, the original mapping assigns each reference opportunity to checkpoint i, and a reply at checkpoint j is eligible when a$\leq$j-i$\leq$b. Native streaming instead uses the continuous difference in audio progress from the reference time, within [a,b] seconds. Each configuration retains the same reference opportunities, observed audio horizon, and emitted replies. Newly eligible pairs are judged under the same content rubric, existing judgments for identical inputs are reused, and order-preserving one-to-one matching is then applied.

Table~\ref{tab:window-sensitivity} includes all 17 simulated and six native configurations. Simulated $\pm1$ checkpoint is exactly the $[-1,+1]$ checkpoint window. Identical numeric bounds across the panels do not imply identical physical-time criteria.

\begin{table}[H]
\centering
\caption{Sensitivity to temporal matching windows for all 23 configurations. Cells report joint Recall/Precision (\%); the default window in each panel is bold.}
\label{tab:window-sensitivity}
\begingroup
\definecolor{panelgray}{HTML}{F1EFF5}
\newsavebox{\pabwindowbox}
\fontsize{9}{10.5}\selectfont
\setlength{\tabcolsep}{1.5pt}
\renewcommand{\arraystretch}{1.02}
\sbox{\pabwindowbox}{%
\begin{tabular}{@{}lrrrrrrr@{}}
\toprule
\rowcolor{panelgray}\multicolumn{8}{@{}l}{\textbf{A. Simulated streaming (checkpoint offsets)}} \\
\addlinespace[2pt]
Model & $[-1,+1]$ & $[-1,+2]$ & $[-1,+3]$ & $[-1,+4]$ & $[-1,+5]$ & $[-2,+2]$ & $[-3,+3]$ \\
\midrule
Qwen3.5-Omni-Plus & \textbf{45.56/45.96} & 45.95/46.36 & 46.14/46.55 & 46.26/46.64 & 46.35/46.74 & 50.31/49.86 & 52.13/51.44 \\
Gemini-3.1-Pro & \textbf{46.92/26.45} & 48.50/27.23 & 49.59/27.71 & 49.94/27.86 & 50.16/27.95 & 52.30/29.00 & 55.88/30.77 \\
Gemini-3.8-Flash & \textbf{34.37/22.18} & 36.57/23.50 & 37.94/24.34 & 38.22/24.46 & 38.38/24.56 & 39.19/25.01 & 42.31/26.88 \\
Gemini-3.7-Flash & \textbf{32.75/18.38} & 34.76/19.47 & 36.66/20.50 & 36.92/20.67 & 36.96/20.69 & 38.17/21.13 & 42.69/23.54 \\
Muse-Spark-1.2 & \textbf{23.03/20.40} & 26.24/23.50 & 28.02/24.48 & 28.67/24.97 & 28.83/25.03 & 28.54/25.30 & 31.32/26.86 \\
Doubao-Seed-2.0-Lite & \textbf{17.23/13.63} & 18.00/14.20 & 18.83/14.85 & 18.93/14.90 & 18.93/14.90 & 21.90/16.35 & 26.53/19.04 \\
MiniCPM-o 4.5 & \textbf{44.93/7.67} & 47.30/8.13 & 48.41/8.39 & 48.60/8.44 & 48.64/8.44 & 50.96/8.88 & 54.29/9.50 \\
Qwen2.5-Omni-7B & \textbf{12.68/9.65} & 12.75/9.70 & 12.86/9.76 & 12.86/9.76 & 12.87/9.80 & 13.71/10.43 & 14.30/10.87 \\
Kimi-Audio-7B & \textbf{28.88/2.90} & 29.76/2.99 & 30.09/3.02 & 30.32/3.05 & 30.36/3.06 & 30.65/3.05 & 31.33/3.11 \\
Baichuan-Omni-1.5 & \textbf{59.30/1.87} & 61.23/1.94 & 62.13/1.97 & 62.32/1.99 & 62.35/1.99 & 62.62/1.98 & 64.11/2.03 \\
Qwen3-Omni-30B-A3B & \textbf{16.53/1.99} & 19.48/2.19 & 22.02/2.32 & 22.31/2.37 & 22.45/2.38 & 21.15/2.37 & 24.81/2.63 \\
Voxtral-Small-24B & \textbf{19.65/1.75} & 20.82/1.88 & 21.79/1.98 & 22.16/2.00 & 22.24/2.01 & 22.09/1.95 & 23.70/2.09 \\
Voxtral-Mini-3B & \textbf{22.93/1.54} & 25.40/1.70 & 26.76/1.79 & 26.94/1.80 & 27.04/1.81 & 26.64/1.76 & 28.42/1.87 \\
MOSS-Audio-8B & \textbf{14.50/1.00} & 14.81/1.02 & 14.92/1.03 & 14.97/1.03 & 15.03/1.03 & 15.02/1.03 & 15.29/1.04 \\
Phi-4-Multimodal & \textbf{1.21/0.82} & 1.21/0.82 & 1.21/0.82 & 1.21/0.82 & 1.21/0.82 & 1.21/0.82 & 1.21/0.82 \\
Qwen2-Audio-7B & \textbf{6.84/0.21} & 7.27/0.22 & 7.37/0.22 & 7.37/0.22 & 7.39/0.22 & 7.94/0.23 & 8.60/0.25 \\
Audio Flamingo Next & \textbf{0.07/0.27} & 0.07/0.27 & 0.07/0.27 & 0.07/0.27 & 0.07/0.27 & 0.07/0.27 & 0.07/0.27 \\
\addlinespace[5pt]
\rowcolor{panelgray}\multicolumn{8}{@{}l}{\textbf{B. Native streaming (seconds of audio progress)}} \\
\addlinespace[2pt]
Model & $[-1,+1]$ & $[-1,+2]$ & $[-1,+3]$ & $[-1,+4]$ & $[-1,+5]$ & $[-2,+2]$ & $[-3,+3]$ \\
\midrule
Qwen3.5-Omni-Plus Realtime & 4.10/9.37 & 7.62/16.68 & \textbf{11.86/32.23} & 15.17/39.42 & 17.41/43.00 & 7.71/16.85 & 12.17/32.71 \\
GPT Realtime 2 & 1.40/3.32 & 4.52/8.81 & \textbf{7.71/12.30} & 11.55/17.08 & 14.00/19.53 & 4.90/9.78 & 8.43/13.66 \\
MiniCPM-o 4.5 & 1.98/4.36 & 4.39/9.05 & \textbf{5.86/12.09} & 7.36/14.37 & 8.54/16.84 & 4.50/9.28 & 5.98/12.32 \\
StepAudio 3 Realtime & 2.16/9.27 & 3.27/11.89 & \textbf{4.54/15.85} & 5.24/17.38 & 5.93/18.60 & 3.61/12.57 & 5.06/16.72 \\
GPT Live 1$^\dagger$ & 2.13/2.24 & 5.57/5.01 & \textbf{11.12/9.52} & 15.94/13.68 & 18.77/16.08 & 6.06/5.22 & 12.16/10.07 \\
Seeduplex$^\dagger$ & 1.45/4.65 & 2.93/8.97 & \textbf{4.97/14.63} & 5.81/17.16 & 7.53/21.03 & 2.93/8.97 & 4.97/14.63 \\
\bottomrule
\end{tabular}}
\ifdim\wd\pabwindowbox>\linewidth
\resizebox{\linewidth}{!}{\usebox{\pabwindowbox}}
\else\usebox{\pabwindowbox}\fi
\endgroup
\par\vspace{3pt}
\begin{minipage}{\linewidth}\footnotesize
$^\dagger$GPT Live 1 and Seeduplex average Detection, Tracking, and Counting because their evaluated interfaces do not support the reference-audio input required for Matching; other configurations average four tasks. Metrics use equal weighting of the three domains within each task. FAR is unchanged across windows because the emitted replies are fixed.
\end{minipage}
\end{table}

\textbf{Sensitivity results.} Wider temporal tolerance admits additional joint matches, with gains varying across protocols and models. All six native configurations benefit more from late tolerance in equal-width comparisons; for example, GPT Realtime 2 reaches 7.71\% Recall under [-1,+3] s versus 4.90\% under [-2,+2] s. In simulated streaming, extending the late boundary from +1 to +3 checkpoints increases Recall from 45.56\% to 46.14\% for Qwen3.5-Omni-Plus and from 46.92\% to 49.59\% for Gemini-3.1-Pro, while some other configurations gain more. The defaults balance tolerance for adjacent-step variation or short initiation waits against credit for substantially early or late notifications; +3 s is not a shared saturation point across models.

\textbf{Response-triggering settings.} Recorded session acknowledgments confirm server VAD in these two interfaces. Even when the listening condition is satisfied within a speech segment, reply initiation may wait until a speech endpoint is detected, making VAD one possible contributor to late offsets. This describes the triggering configuration used in our runs; the comparison does not isolate the effect of VAD or establish a common delay mechanism for all native systems. Settings are summarized in Table~\ref{tab:vad-settings}.

\begin{table}[htbp]
\centering
\caption{Response-triggering settings confirmed by session acknowledgments.}
\label{tab:vad-settings}
\small\setlength{\tabcolsep}{6pt}
\begin{tabular}{@{}llr@{}}
\toprule
Model & Trigger & Silence duration \\
\midrule
GPT Realtime 2 & Server VAD & 500 ms \\
Qwen3.5-Omni-Plus Realtime & Server VAD & 800 ms \\
\bottomrule
\end{tabular}
\end{table}

Both interfaces use a VAD threshold of 0.5, 300\,ms prefix padding, automatic response creation, and disabled response interruption; audio is sent in 200\,ms chunks.

\textbf{Interpreting distributions.} Timing histograms use positive requests. Correct outputs use their matched reference; unmatched outputs use the nearest reference only for visualization. Each class is normalized within each model before averaging across models. Correct-output distributions are conditioned on the matching window, while simulated late-side comparisons are limited by the observed audio horizon; unobserved later replies cannot be inferred. The sweep changes scoring tolerance on fixed traces, not response behavior.

\subsection{Joint Event F1}
\label{app:f1-definition}

We report macro-averaged joint event F1 as a summary measure while retaining Recall, Precision, and negative-request FAR to characterize notification behavior. Let $t$ index the task (Detection, Tracking, Counting, or Matching) and $d$ the audio domain (sound, speech, or music). Within each task--domain cell, $\mathrm{TP}_{t,d}$ counts one-to-one matches satisfying both timing and content requirements, $M_{t,d}$ counts reference notification opportunities, and $K_{t,d}$ counts emitted non-control replies, including incorrect, repeated, and negative-request outputs. We define
\begin{equation}
F_{1,t,d}=\frac{2\mathrm{TP}_{t,d}}{M_{t,d}+K_{t,d}}.
\label{eq:joint-event-f1}
\end{equation}
With $\mathrm{FP}_{t,d}=K_{t,d}-\mathrm{TP}_{t,d}$ and $\mathrm{FN}_{t,d}=M_{t,d}-\mathrm{TP}_{t,d}$, this is equivalent to $2\mathrm{TP}/(2\mathrm{TP}+\mathrm{FP}+\mathrm{FN})$. We average equally over the three domains within each task and then over the four tasks:
\begin{equation}
F_1^{\mathrm{overall}}=\frac{1}{4}\sum_{t\in\mathcal{T}}\frac{1}{3}\sum_{d\in\mathcal{D}} F_{1,t,d},
\label{eq:macro-joint-event-f1}
\end{equation}
where $\mathcal{T}$ and $\mathcal{D}$ denote the task and domain sets. Timing and content requirements follow the existing matching rules. The reported Recall retains its positive-request averaging, whereas F1 is computed from event counts. Because Recall averages requests while F1 pools event counts within each task--domain cell, F1 is not the harmonic mean of the displayed macro Recall and Precision. Table~\ref{tab:joint-event-f1} reports the complete results.

\clearpage
\section{Complete Benchmark Results}
\label{app:additional-results}
\label{app:coverage-notes}

\subsection{Coverage and Task-Level Results}
Tables~\ref{tab:full-results} and~\ref{tab:full-results-cm} report all 23 model--interface configurations, split by task to keep the numbers readable. The main table presents 13 configurations. Qwen Realtime denotes Qwen3.5-Omni-Plus Realtime. The simulated open-source models in the main table are Qwen3-Omni~\citep{qwen3omni2025}, MiniCPM-o 4.5~\citep{arxiv2604_27393}, Kimi-Audio~\citep{arxiv2504_18425}, and Baichuan-Omni-1.5~\citep{arxiv2501_15368}.

The tables retain the evaluated request count $n$ for every configuration. Matching is unavailable for Seeduplex and GPT Live 1 under the evaluated reference-audio interfaces. Their restricted averages are reported separately from four-task averages in Table~\ref{tab:joint-event-f1}. GPT Live 1 ($^\ddagger$) uses reconstructed output episodes. Evaluation procedures, including response reconstruction and observation horizons, appear in Appendix~\ref{app:evaluation-details}.

\begin{table}[H]
\centering
\caption{Complete Detection and Tracking results (\%). R/P/FAR denote joint recall, joint precision, and negative-request false-alarm rate.}
\label{tab:full-results}
\begingroup\fontsize{9.5}{11}\selectfont
\setlength{\tabcolsep}{5pt}\renewcommand{\arraystretch}{1.12}
\begin{tabular*}{\linewidth}{@{\extracolsep{\fill}}lr*{6}{r}@{}}
\toprule
 & & \multicolumn{3}{c}{Detection} & \multicolumn{3}{c}{Tracking} \\
\cmidrule(lr){3-5}\cmidrule(lr){6-8}
Model & $n$ & R$\uparrow$ & P$\uparrow$ & FAR$\downarrow$ & R$\uparrow$ & P$\uparrow$ & FAR$\downarrow$ \\
\midrule
\multicolumn{8}{@{}l}{\textbf{Simulated streaming: open-source}} \\
Phi-4-Multimodal & 1999 & $1.7$ & $1.7$ & $3.4$ & $3.1$ & $1.5$ & $4.5$ \\
Audio Flamingo Next & 1999 & $0.0$ & $0.0$ & $1.5$ & $0.3$ & $1.1$ & $0.9$ \\
Voxtral-Mini-3B & 1999 & $22.9$ & $1.0$ & $99.6$ & $37.4$ & $4.4$ & $99.5$ \\
Voxtral-Small-24B & 1999 & $19.7$ & $1.3$ & $62.7$ & $31.8$ & $4.8$ & $75.7$ \\
MOSS-Audio-8B & 1999 & $12.0$ & $0.7$ & $64.0$ & $18.2$ & $2.8$ & $86.0$ \\
Baichuan-Omni-1.5 & 1999 & $58.3$ & $1.4$ & $98.0$ & $71.2$ & $4.5$ & $99.5$ \\
Qwen2-Audio-7B & 1999 & $8.2$ & $0.2$ & $93.2$ & $10.6$ & $0.5$ & $94.6$ \\
Qwen2.5-Omni-7B & 1999 & $20.8$ & $13.0$ & $29.9$ & $18.5$ & $18.2$ & $39.6$ \\
Qwen3-Omni-30B-A3B & 1999 & $16.6$ & $1.9$ & $69.4$ & $24.8$ & $3.7$ & $81.5$ \\
Kimi-Audio-7B & 1999 & $25.4$ & $1.9$ & $52.7$ & $42.4$ & $8.2$ & $65.8$ \\
MiniCPM-o 4.5 & 1999 & $50.3$ & $9.1$ & $93.4$ & $67.5$ & $14.1$ & $95.5$ \\
\midrule
\multicolumn{8}{@{}l}{\textbf{Simulated streaming: closed-source}} \\
Gemini-3.1-Pro & 1999 & $48.0$ & $25.3$ & $53.1$ & $75.8$ & $43.9$ & $59.9$ \\
Gemini-3.7-Flash & 1996 & $26.4$ & $11.0$ & $84.3$ & $57.8$ & $35.4$ & $78.4$ \\
Gemini-3.8-Flash & 1993 & $28.3$ & $13.5$ & $72.8$ & $53.3$ & $38.6$ & $76.6$ \\
Qwen3.5-Omni-Plus & 1990 & $52.1$ & $57.7$ & $14.4$ & $68.8$ & $65.2$ & $27.9$ \\
Muse-Spark-1.2 & 1999 & $26.0$ & $30.1$ & $27.3$ & $40.1$ & $32.1$ & $50.5$ \\
Doubao-Seed-2.0-Lite & 1999 & $27.9$ & $14.1$ & $71.8$ & $25.8$ & $31.7$ & $77.0$ \\
\midrule
\multicolumn{8}{@{}l}{\textbf{Native streaming}} \\
MiniCPM-o 4.5 & 1999 & $9.0$ & $13.0$ & $15.4$ & $4.0$ & $22.6$ & $39.6$ \\
Qwen3.5-Omni-Plus Realtime & 1999 & $21.5$ & $35.3$ & $4.0$ & $15.8$ & $45.6$ & $9.5$ \\
GPT Realtime 2 & 1998 & $12.3$ & $10.1$ & $12.6$ & $11.1$ & $27.2$ & $17.6$ \\
StepAudio 3 Realtime & 1999 & $8.3$ & $12.5$ & $5.3$ & $3.9$ & $21.4$ & $0.9$ \\
Seeduplex & 1483 & $6.5$ & $8.9$ & $3.8$ & $6.5$ & $29.4$ & $3.2$ \\
GPT Live 1$^\ddagger$ & 1534 & $13.6$ & $7.2$ & $23.9$ & $18.1$ & $20.9$ & $41.8$ \\
\bottomrule
\end{tabular*}\endgroup
\end{table}

\begin{table}[H]
\centering
\caption{Complete Counting and Matching results (\%). R/P/FAR denote joint recall, joint precision, and negative-request false-alarm rate.}
\label{tab:full-results-cm}
\begingroup\fontsize{9.5}{11}\selectfont
\setlength{\tabcolsep}{5pt}\renewcommand{\arraystretch}{1.12}
\begin{tabular*}{\linewidth}{@{\extracolsep{\fill}}lr*{6}{r}@{}}
\toprule
 & & \multicolumn{3}{c}{Counting} & \multicolumn{3}{c}{Matching} \\
\cmidrule(lr){3-5}\cmidrule(lr){6-8}
Model & $n$ & R$\uparrow$ & P$\uparrow$ & FAR$\downarrow$ & R$\uparrow$ & P$\uparrow$ & FAR$\downarrow$ \\
\midrule
\multicolumn{8}{@{}l}{\textbf{Simulated streaming: open-source}} \\
Phi-4-Multimodal & 1999 & $0.0$ & $0.0$ & $0.5$ & $0.0$ & $0.0$ & $0.0$ \\
Audio Flamingo Next & 1999 & $0.0$ & $0.0$ & $1.1$ & $0.0$ & $0.0$ & $0.5$ \\
Voxtral-Mini-3B & 1999 & $13.1$ & $0.5$ & $99.1$ & $18.3$ & $0.3$ & $99.6$ \\
Voxtral-Small-24B & 1999 & $14.5$ & $0.6$ & $81.4$ & $12.5$ & $0.4$ & $74.2$ \\
MOSS-Audio-8B & 1999 & $22.5$ & $0.5$ & $88.6$ & $5.2$ & $0.1$ & $89.9$ \\
Baichuan-Omni-1.5 & 1999 & $59.4$ & $0.9$ & $100.0$ & $48.3$ & $0.7$ & $100.0$ \\
Qwen2-Audio-7B & 1999 & $6.9$ & $0.1$ & $92.4$ & $1.7$ & $0.1$ & $81.3$ \\
Qwen2.5-Omni-7B & 1999 & $4.4$ & $6.0$ & $14.6$ & $7.0$ & $1.5$ & $47.0$ \\
Qwen3-Omni-30B-A3B & 1999 & $10.9$ & $0.4$ & $89.0$ & $13.8$ & $1.9$ & $73.4$ \\
Kimi-Audio-7B & 1999 & $29.7$ & $0.8$ & $62.4$ & $17.9$ & $0.7$ & $57.8$ \\
MiniCPM-o 4.5 & 1999 & $29.7$ & $2.2$ & $96.4$ & $32.2$ & $5.3$ & $93.8$ \\
\midrule
\multicolumn{8}{@{}l}{\textbf{Simulated streaming: closed-source}} \\
Gemini-3.1-Pro & 1999 & $29.4$ & $20.0$ & $68.5$ & $34.4$ & $16.6$ & $84.2$ \\
Gemini-3.7-Flash & 1996 & $19.5$ & $10.1$ & $68.0$ & $27.2$ & $17.1$ & $58.7$ \\
Gemini-3.8-Flash & 1993 & $20.2$ & $11.4$ & $58.9$ & $35.8$ & $25.3$ & $46.7$ \\
Qwen3.5-Omni-Plus & 1990 & $15.9$ & $23.2$ & $26.2$ & $45.5$ & $37.8$ & $30.0$ \\
Muse-Spark-1.2 & 1999 & $7.5$ & $5.8$ & $82.4$ & $18.5$ & $13.6$ & $76.9$ \\
Doubao-Seed-2.0-Lite & 1999 & $5.8$ & $3.7$ & $82.2$ & $9.5$ & $5.0$ & $98.6$ \\
\midrule
\multicolumn{8}{@{}l}{\textbf{Native streaming}} \\
MiniCPM-o 4.5 & 1999 & $5.2$ & $7.4$ & $53.6$ & $5.2$ & $5.4$ & $51.1$ \\
Qwen3.5-Omni-Plus Realtime & 1999 & $1.7$ & $4.0$ & $13.7$ & $8.4$ & $44.1$ & $14.1$ \\
GPT Realtime 2 & 1998 & $3.6$ & $5.6$ & $36.3$ & $3.8$ & $6.3$ & $25.8$ \\
StepAudio 3 Realtime & 1999 & $0.8$ & $17.1$ & $8.5$ & $5.1$ & $12.5$ & $9.8$ \\
Seeduplex & 1483 & $1.9$ & $5.6$ & $2.4$ & --- & --- & --- \\
GPT Live 1$^\ddagger$ & 1534 & $1.6$ & $0.5$ & $45.4$ & --- & --- & --- \\
\bottomrule
\end{tabular*}\endgroup
\end{table}

\clearpage
\subsection{Overall Results and Joint Event F1}
\label{app:joint-event-f1}
Table~\ref{tab:joint-event-f1} reports the overall metrics. F1 is computed from event counts within each task--domain cell before macro-averaging, as defined in Appendix~\ref{app:f1-definition}. The two native interfaces without Matching are shown in a separate three-task block.

\begin{table}[H]
\centering
\caption{Joint event F1 and accompanying metrics for 23 model--interface configurations (\%).}
\label{tab:joint-event-f1}
\begingroup
\definecolor{pabf1shade}{HTML}{F0ECF7}
\definecolor{pabf1group}{HTML}{F1F2F4}
\setlength{\tabcolsep}{5pt}
\renewcommand{\arraystretch}{1.06}
\fontsize{9.5}{11}\selectfont
\begin{tabularx}{\linewidth}{@{}p{.35\linewidth}>{\raggedleft\arraybackslash}p{.07\linewidth}*{2}{>{\raggedleft\arraybackslash}X}>{\columncolor{pabf1shade}\raggedleft\arraybackslash}X>{\raggedleft\arraybackslash}X@{}}
\toprule
\textbf{Model} & $n$ & \textbf{Recall}$\uparrow$ & \textbf{Precision}$\uparrow$ & \textbf{F1}$\uparrow$ & \textbf{FAR}$\downarrow$ \\
\midrule
\rowcolor{pabf1group}
\multicolumn{6}{@{}l@{}}{\hspace{4pt}\strut\textbf{(a) Simulated streaming}} \\
\addlinespace[2pt]
Qwen3.5-Omni-Plus & 1990 & 45.56 & 45.96 & \textbf{44.41} & 24.62 \\
Gemini-3.1-Pro & 1999 & 46.92 & 26.45 & 32.83 & 66.42 \\
Gemini-3.8-Flash & 1993 & 34.37 & 22.18 & 26.08 & 63.76 \\
Gemini-3.7-Flash & 1996 & 32.75 & 18.38 & 22.90 & 72.37 \\
Muse-Spark-1.2 & 1999 & 23.03 & 20.40 & 17.39 & 59.26 \\
Doubao-Seed-2.0-Lite & 1999 & 17.23 & 13.63 & 13.35 & 82.41 \\
MiniCPM-o 4.5 & 1999 & 44.93 & 7.67 & 11.78 & 94.78 \\
Qwen2.5-Omni-7B & 1999 & 12.68 & 9.65 & 8.85 & 32.80 \\
Kimi-Audio-7B & 1999 & 28.88 & 2.90 & 4.47 & 59.69 \\
Baichuan-Omni-1.5 & 1999 & 59.30 & 1.87 & 3.46 & 99.39 \\
Qwen3-Omni-30B-A3B & 1999 & 16.53 & 1.99 & 3.14 & 78.32 \\
Voxtral-Small-24B & 1999 & 19.65 & 1.75 & 3.01 & 73.48 \\
Voxtral-Mini-3B & 1999 & 22.93 & 1.54 & 2.65 & 99.44 \\
MOSS-Audio-8B & 1999 & 14.50 & 1.00 & 1.70 & 82.13 \\
Phi-4-Multimodal & 1999 & 1.21 & 0.82 & 0.70 & 2.09 \\
Qwen2-Audio-7B & 1999 & 6.84 & 0.21 & 0.39 & 90.38 \\
Audio Flamingo Next & 1999 & 0.07 & 0.27 & 0.10 & 0.98 \\
\midrule
\rowcolor{pabf1group}
\multicolumn{6}{@{}l@{}}{\hspace{4pt}\strut\textbf{(b) Native streaming}} \\
\addlinespace[2pt]
Qwen3.5-Omni-Plus Realtime & 1999 & 11.86 & 32.23 & \textbf{13.76} & 10.32 \\
GPT Realtime 2 & 1998 & 7.71 & 12.30 & 6.85 & 23.05 \\
MiniCPM-o 4.5 & 1999 & 5.86 & 12.09 & 6.67 & 39.93 \\
StepAudio 3 Realtime & 1999 & 4.54 & 15.85 & 5.44 & 6.11 \\
\midrule
\multicolumn{6}{@{}l}{\textbf{Native streaming: three-task means (Matching unavailable)}} \\
GPT Live 1$^{\dagger}$ & 1534 & 11.12 & 9.52 & 7.47 & 37.03 \\
Seeduplex$^{\dagger}$ & 1483 & 4.97 & 14.63 & 5.51 & 3.12 \\
\bottomrule
\end{tabularx}
\par\vspace{5pt}
\begin{minipage}{\linewidth}
\fontsize{8.5}{10}\selectfont\raggedright
$n$ denotes scoreable requests. $^{\dagger}$Due to reference-audio input limitations in the evaluated interfaces, GPT Live 1 and Seeduplex are evaluated only on Detection, Tracking, and Counting. Their scores average these three tasks; other models average all four.
\end{minipage}
\endgroup
\end{table}

Qwen3.5-Omni-Plus leads simulated-streaming F1 (44.41\%); Qwen Realtime leads the four-task native results (13.76\%). Baichuan-Omni reaches 59.30\% recall but only 3.46\% F1, illustrating the cost of excess output.

\clearpage
\subsection{Supplementary Analysis of Timing and Content Constraints}
\label{app:content-ablation}

We hold requests, references, outputs, observation horizons, and temporal windows fixed. Time-only matching removes content acceptance while retaining the same order-preserving assignment and recall aggregation (Section~\ref{sec:evaluation}). Table~\ref{tab:timing-content-ablation} compares it with joint matching.

\begin{table}[H]
\centering
\caption{Time-only and joint recall (\%), with identical outputs and timing windows.}
\label{tab:timing-content-ablation}
\small
\setlength{\tabcolsep}{4pt}
\begin{tabular}{lrrrr}
\toprule
Model & Mode & Time-only R & Joint R & Gap (pp) \\
\midrule
Baichuan-Omni & Simulated & 93.6 & 59.3 & 34.3 \\
Qwen3-Omni & Simulated & 47.1 & 16.5 & 30.5 \\
Kimi-Audio & Simulated & 50.5 & 28.9 & 21.7 \\
MiniCPM-o 4.5 & Simulated & 52.4 & 44.9 & 7.5 \\
Gemini-3.1-Pro & Simulated & 50.0 & 46.9 & 3.1 \\
Gemini-3.7-Flash & Simulated & 35.1 & 32.8 & 2.3 \\
Qwen3.5-Omni-Plus & Simulated & 47.0 & 45.6 & 1.4 \\
Muse-Spark-1.2 & Simulated & 25.3 & 23.0 & 2.3 \\
Seed 2.0 Lite & Simulated & 18.3 & 17.2 & 1.1 \\
MiniCPM-o 4.5 & Native & 7.7 & 5.9 & 1.9 \\
Qwen Realtime & Native & 12.1 & 11.9 & 0.2 \\
GPT Realtime 2 & Native & 9.4 & 7.7 & 1.7 \\
StepAudio 3 Realtime & Native & 5.5 & 4.5 & 1.0 \\
\bottomrule
\end{tabular}
\end{table}

Gaps are percentage points computed before rounding, not per-reply content error rates. Optimal assignments can change after content filtering. Time-only matches may include acknowledgments, transcriptions, or unrelated replies and therefore do not establish correct target perception.

\clearpage
\subsection{Performance across Audio Domains and Tasks}
\label{app:domain-task-heatmaps}

\begin{figure}[H]
\centering
\includegraphics[width=\linewidth]{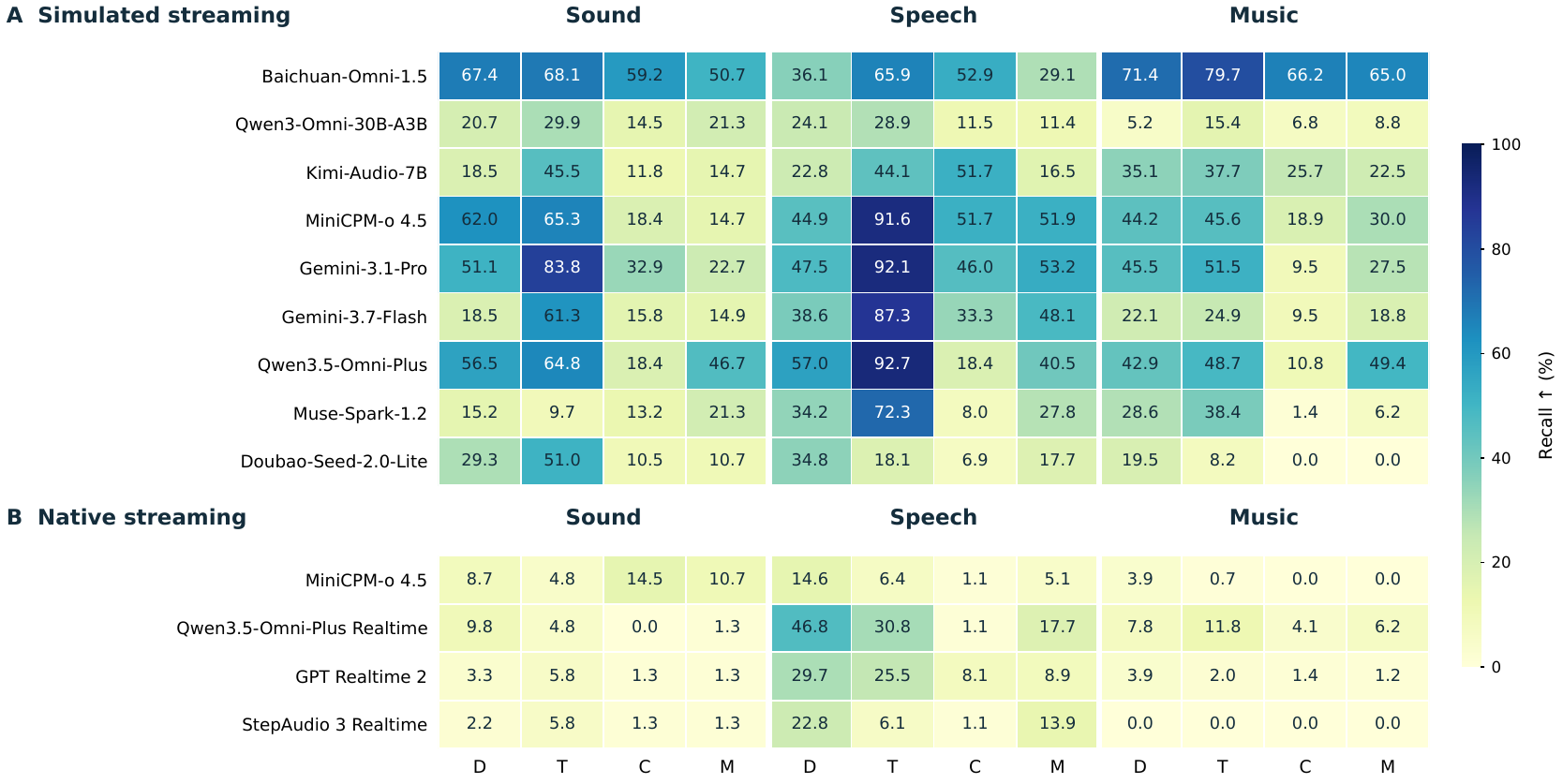}
\caption{Recall by audio domain and task (\%, higher is better).
D/T/C/M denote Detection/Tracking/Counting/Matching. All heatmaps share the
same ordering; cells show percentages to one decimal place.}
\label{fig:domain-recall}
\end{figure}

\begin{figure}[H]
\centering
\includegraphics[width=\linewidth]{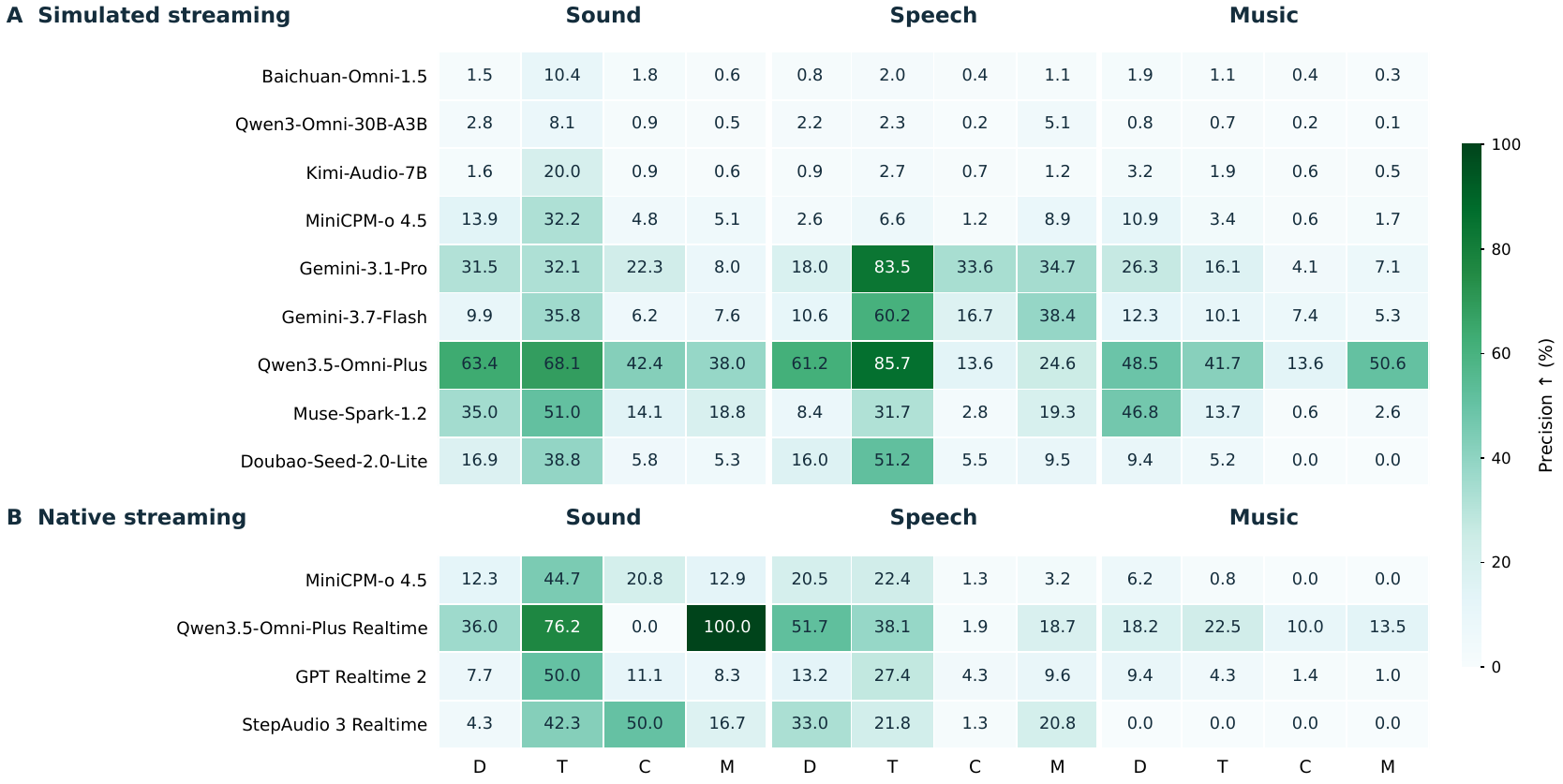}
\caption{Precision by audio domain and task (\%, higher is better), pooling
notifications from positive and negative requests. Layout follows
Figure~\ref{fig:domain-recall}.}
\label{fig:domain-precision}
\end{figure}

\begin{figure}[H]
\centering
\includegraphics[width=\linewidth]{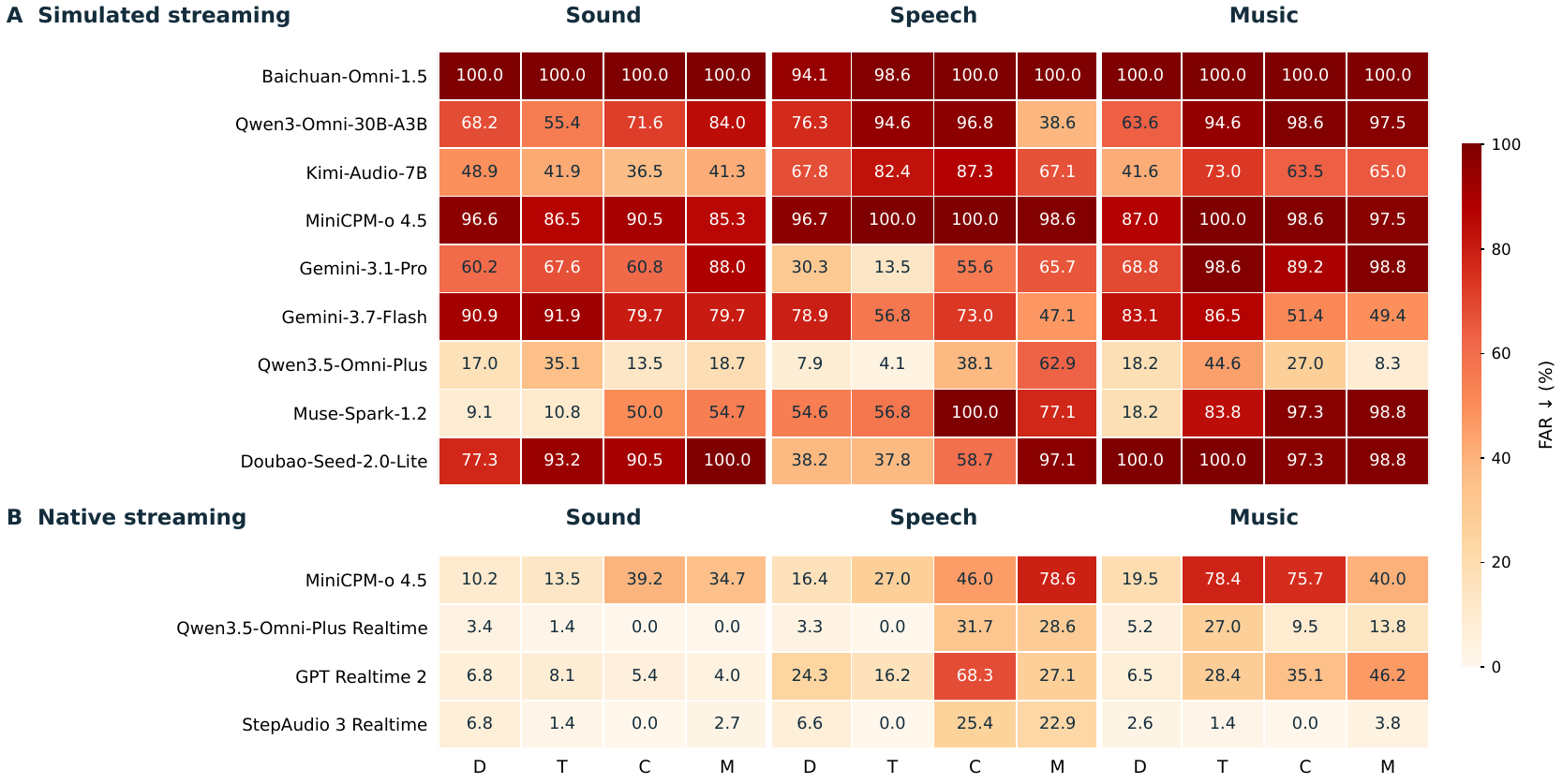}
\caption{FAR by audio domain and task (\%, lower is better): the percentage
of negative requests with any notification. Darker cells indicate more false alarms.}
\label{fig:domain-far}
\end{figure}

Figures~\ref{fig:domain-recall}--\ref{fig:domain-far} show the 13 main-table configurations across the 12 domain--task combinations, using the same model order. Each cell uses the metric definitions in Section~\ref{sec:evaluation}; main-table task scores average the three domains equally. Results are compared within each streaming mode.

\textbf{Task-dependent domain strengths.} Qwen3.5-Omni-Plus reaches 64.8\%/92.7\%/48.7\% Tracking recall on Sound/Speech/Music, but Music exceeds Speech in Matching (49.4\% versus 40.5\%). Domain rankings therefore depend on the listening task.

\textbf{Complementary metrics.} On Speech Tracking, simulated MiniCPM-o 4.5 and Qwen have similar recall (91.6\%/92.7\%) but precision of 6.6\%/85.7\% and FAR of 100.0\%/4.1\%, emitting 5,564/428 replies over 150 requests. Native Qwen has zero FAR but no output on Sound Counting; its 100\% Sound Matching precision reflects one correct reply and only 1.3\% recall. Neither recall, precision, nor FAR alone captures reliable listening. These subset comparisons do not isolate underlying cognitive mechanisms.

\clearpage

\clearpage
\section{Capability Diagnostics}
\label{app:capability-diagnostics}
\begingroup
\definecolor{streambg}{RGB}{236,242,248}
\definecolor{avgbg}{RGB}{242,246,249}

\subsection{Factual QA and Listening-Instruction Responses}

The experiment uses 112 requests, with 28 per task: 16 positive and 12 negative. Condition A reformulates the listening request as a factual question about target presence, reference matching, or whether the specified occurrence has been reached. Condition B retains the original listening instruction. Both conditions use the same audio prefix, reference audio, neutral system prompt, and fixed model configuration, independently generating one reply.

A scores the factual answer; B scores the notification decision and required content. Both are binary one-shot accuracies, averaged equally across domains. B is distinct from streaming complete-request success.

Table~\ref{tab:task-ab-probes} in the main text reports the four models with results across the core counting and localization probes. Supplementary A/B results for two additional models are shown below (\%).

\begin{table}[H]
\centering
\caption{Additional factual QA / listening-response accuracies (A/B, \%). Both conditions use one-shot offline inputs.}
\label{tab:diagnostic-ab-extra}
\begingroup
\small
\setlength{\tabcolsep}{5pt}
\renewcommand{\arraystretch}{1.12}
\begin{tabular}{@{}lrrrr@{}}
\toprule
Model & Detection & Tracking & Counting & Matching \\
\midrule
Qwen2-Audio & 35.9/21.5 & 35.8/18.3 & 30.8/34.2 & 16.7/24.2 \\
Qwen2.5-Omni & 75.2/43.0 & 64.2/39.2 & 15.0/21.7 & 44.2/25.0 \\
\bottomrule
\end{tabular}
\endgroup

\end{table}

For the four main-text models, A outperforms B in 15 of 16 model–task comparisons; MiniCPM-o 4.5 on Counting is the exception. This comparison supports a performance gap between factual judgment and execution of listening requests, but does not isolate internal perception, memory, or decision mechanisms.

\subsection{Counting under Different Conditions}

The counting comparison uses the same 28 requests: eight Sound, ten Speech, and ten Music, comprising 16 positives and 12 negatives. The first three conditions share the same audio prefix and independently elicit one response: retaining the original listening instruction, asking whether $N$ has been reached, or asking for the actual total number of target occurrences. The final column uses the original simulated-streaming traces for the same requests.

The streaming comparison uses the same requests but longer observations for all 16 positives, different generation budgets and success criteria, and different time anchors for six Speech requests. It is therefore descriptive, not a wording-only intervention or a causal estimate of the cost of proactive notification.

\begin{table}[H]
\centering
\caption{Counting performance under four conditions (\%). The first three columns use identical one-shot audio prefixes; the final column reports simulated-streaming complete-request success.}
\label{tab:diagnostic-counting}
\begingroup
\definecolor{streambg}{RGB}{236,242,248}
\fontsize{9.5}{11}\selectfont
\setlength{\tabcolsep}{3pt}
\renewcommand{\arraystretch}{1.12}
\begin{tabular*}{\linewidth}{@{\extracolsep{\fill}}lrrr>{\columncolor{streambg}}r@{}}
\toprule
& \multicolumn{3}{c}{One-shot accuracy} & \multicolumn{1}{>{\columncolor{streambg}}c}{Sim. stream} \\
\cmidrule(lr){2-4}
Model & Listen (B) & Reached $N$ (A) & Exact count & Complete success \\
\midrule
Qwen3-Omni & 15.0 & 42.5 & 15.8 & 11.7 \\
MiniCPM-o 4.5 & 50.8 & 45.0 & 21.7 & 4.2 \\
Baichuan-Omni & 39.2 & 50.8 & 31.7 & 0.0 \\
Seed 2.0 Lite & 33.3 & 74.2 & 56.7 & 14.2 \\
\bottomrule
\end{tabular*}
\endgroup

\end{table}

All four measures average request-level accuracy or success equally across domains. Listen and Reached $N$ correspond to Counting B and A in Table~\ref{tab:task-ab-probes}.

Threshold QA asks whether the requested occurrence has been reached, not whether the total count is exactly $N$. Exact counting requires the answer to match the actual number of occurrences in the supplied audio. The correct total is not provided in the question. Simulated-streaming success requires all reference opportunities to be correctly matched without extra replies; negative requests require silence throughout. An always-yes threshold baseline scores 56.7\%, while an always-silent streaming baseline scores 43.3\% complete-request success.

\par\noindent\begin{minipage}{\linewidth}
\textbf{Triangle-strike example.} The instruction requests the third triangle strike, but the supplied audio contains two. Seed correctly answers both counting questions, yet its one-shot listening response says, “The third triangle strike has occurred.” A separate streaming trace makes the same claim at three seconds (translated). Correct factual counting does not guarantee correct instruction execution.
\end{minipage}

\subsection{All-Occurrence Interval Localization}

Table~\ref{tab:counting-grounding} summarizes the domain-averaged results in the main text. The grounding probe uses a separate set of 60 positive recordings with complete interval annotations: 31 Sound, 17 Speech, and 12 Music. Each model receives a complete recording once and reports all target onset–offset intervals, without choosing when to notify during continuous listening. We calculate temporal IoU between the unions of predicted and reference intervals per recording, average within each domain, and then average equally across domains.

\begin{table}[H]
\centering
\caption{Interval localization by domain (temporal union mIoU $\times100$). Avg. weights the three domains equally; Full-span always predicts the entire recording.}
\label{tab:diagnostic-localization-domains}
\begingroup
\small
\setlength{\tabcolsep}{5pt}
\renewcommand{\arraystretch}{1.12}
\begin{tabular}{@{}lrrr>{\columncolor{avgbg}}r@{}}
\toprule
Model & Sound & Speech & Music & Avg. \\
\midrule
Qwen3-Omni & 13.4 & 1.8 & 7.5 & 7.6 \\
MiniCPM-o 4.5 & 1.4 & 0.0 & 16.4 & 5.9 \\
Baichuan-Omni & 5.9 & 1.6 & 23.9 & 10.5 \\
Seed 2.0 Lite & 34.3 & 40.3 & 69.2 & 47.9 \\
\midrule
\textit{Full-span} & 9.3 & 4.8 & 73.3 & 29.1 \\
\bottomrule
\end{tabular}
\endgroup

\end{table}

Full-span predicts the entire recording. Negative or out-of-recording timestamps count as false predicted time; reversed, non-finite, ambiguous, or unparseable predictions score zero.

All recordings are target-present. Speech targets occur once; Music targets are continuous vocal regions. The set therefore does not test negative detection, repeated-word localization, or all music-event types. Full-span scores 73.3 on Music, above Seed's 69.2, showing that long targets can reward coarse predictions.

\subsection{Instruction Placement and Repetition}

On the same 112 requests, we compare three input arrangements across six models: a prepended instruction, the same instruction before and after the audio, and an appended instruction only. The main audio, applicable reference audio, system prompt, and generation settings are held fixed. Each condition independently receives one input and generates one reply. Repetition supplies no additional event times, actual counts, or answers about whether to notify.

We use one-shot listening-response accuracy, averaging equally across domains and then tasks (\%).

\begin{table}[H]
\centering
\caption{One-shot listening-response accuracy under different instruction placements (\%), averaged equally across domains and tasks.}
\label{tab:diagnostic-instruction-position}
\begingroup
\small
\setlength{\tabcolsep}{5pt}
\renewcommand{\arraystretch}{1.12}
\begin{tabular}{@{}lrrr@{}}
\toprule
Model & Prepended & Repeated & Appended \\
\midrule
Qwen2-Audio & 7.3 & 23.5 & 24.5 \\
Qwen2.5-Omni & 28.7 & 32.7 & 32.2 \\
Qwen3-Omni & 41.4 & 35.2 & 27.0 \\
MiniCPM-o 4.5 & 41.9 & 44.8 & 40.6 \\
Baichuan-Omni & 23.1 & 25.3 & 29.0 \\
Seed 2.0 Lite & 52.7 & 51.6 & 51.7 \\
\bottomrule
\end{tabular}
\endgroup

\end{table}

Repetition improves Qwen2-Audio from 7.3\% to 23.5\%, but appending alone reaches 24.5\%; Qwen3-Omni decreases from 41.4\% to 35.2\%. Within Qwen2.5-Omni, Detection improves from 28.9\% to 43.3\% while Matching falls from 35.8\% to 25.0\%. Instruction placement has model- and task-dependent effects, and this one-shot probe does not independently establish streaming memory capability.
\endgroup

\clearpage
\section{Controlled Notification Rules: Full Results}
\label{app:rule-heatmaps}
\label{sec:rule-contrast}

\subsection{Paired Audio and Notification Rules}
The set contains 25 verified audio--target groups (eight Sound, 13 Speech, four Music), each at most 90 seconds long. First, Every, and Nth-only ($N=2$) use identical audio, targets, and full-recording observation: 75 requests per configuration and 525 traces overall. Simulated streaming uses one-second increments, actual reply history, and $\pm1$ checkpoint matching; native streaming uses $[-1,+3]$ seconds of audio progress. Results are separated by mode. MiniCPM receives the condition through system context. All requests are positive, so negative-request FAR is undefined.

\subsection{Reading the Rule-Comparison Plots}
Columns index verified target occurrences, regardless of the ordinal claimed in a reply. Blue counts occurrences with non-control output in their timing windows; repeated output within a window counts once. Green preserves the order-preserving one-to-one matches satisfying the rule, timing, and content criteria. Blue includes green. Outlines mark required positions. Green zeros elsewhere are imposed by the rule, so blue must be consulted to assess whether the model spoke.

\textbf{Denominators and omitted outputs.}
The first three columns use all 25 groups; 4th+ pools 41 later occurrences from 16 groups. Labels show numerator/denominator and colors show the ratio, not domain-macro recall: Seed First is 15/25 but 50.64\% after domain averaging; Nth-only is 1/25 but 2.56\%. Off-window counts requests with output outside every target window; Silent counts requests with no output. Both denominators are 25. In-window and off-window replies can coexist in a request, so rows need not sum to 25.

Two of 2,167 outputs overlap adjacent windows, both from Qwen Realtime on one group under First and Every. Blue counts both covered windows; green retains the unique joint assignment.

\subsection{Results and Paired Examples}
Seed correctly reports the first three Every occurrences in 15, seven, and two of the same 25 groups, compared with 15, 18, and 17 for Gemini-3.7-Flash. Seed also has output near the second and third occurrences in only seven and two groups, so its decline is visible before content filtering. Conversely, Muse covers these occurrences in 11, eight, and 11 groups but correctly matches only six, three, and three. The plots distinguish missing timely output from outputs that fail content or rule requirements.

In native streaming, Qwen Realtime has no Nth-only output near the first occurrence, but 12/25 requests are silent, 9/25 include off-window output, and only 3/25 correctly notify the second occurrence. StepAudio is silent on 19/25 Nth-only requests, and all seven remaining outputs are off-window. Low early-event output alone does not establish correct waiting.

\textbf{Checking the occurrence claims.}
All 25 Seed Nth-only requests produce one reply, each explicitly claiming a second occurrence. Nine precede the first eligible window, 11 fall near the first occurrence, and three lie between the first and second windows. One correctly matches the second occurrence; one follows its window and falls outside every target window. Thus, 23/25 replies precede the second eligible window. This content-specific finding is limited to the inspected Seed replies; other models' outputs can include acknowledgments.

For the ``county town'' example, the first two source occurrences begin at 3.12 and 12.00 seconds. Seed responds at 4 seconds under both First and Nth-only, changing only its occurrence claim. Identical timing is not universal: among 24 groups with exactly one Seed reply under both rules, only five have the same reply time.

\textbf{Paired checks and interpretation.}
Among the 15 groups with a valid Seed First notification, Nth-only responds near the first occurrence in nine and correctly notifies the second in none. Gemini and Seed both correctly report the first Every occurrence on 11 groups (two Sound and nine Speech); they jointly match 30 and six of the 33 subsequent opportunities, respectively. This conditional subset supplements the full results and does not extend to Music. The study controls audio and observation horizon, but does not isolate perception, event segmentation, memory, or counting mechanisms.

Correct-notification counts measure opportunity-level joint hits; complete-request success also excludes extra replies, explaining Seed's 15/25 versus 14/25 First results. Ordinal 1--11 denominators are 25, 25, 25, 16, 8, 5, 5, 3, 2, 1, and 1; the sparse tail does not support a stable count trend.

\clearpage
\begin{figure}[!ht]
\centering
\includegraphics[width=\linewidth]{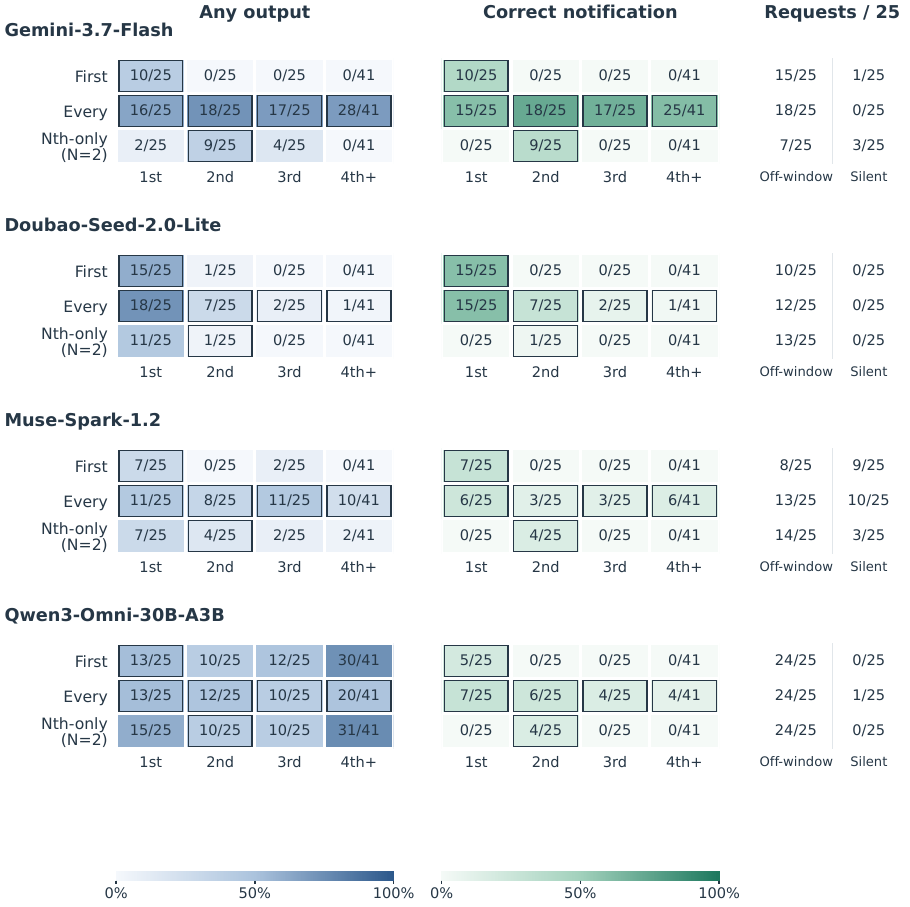}
\caption{Full rule contrasts for four simulated configurations. Blue shows any-output coverage; green shows joint correctness. Outlines mark required positions. The first three columns each use 25 groups; 4th+ pools 41 later occurrences. Off-window and Silent count requests out of 25. Colors encode coverage fractions.}
\label{fig:rule-appendix-simulated}\label{fig:rule-contrast}
\end{figure}

\clearpage
\begin{figure}[!ht]
\centering
\includegraphics[width=\linewidth]{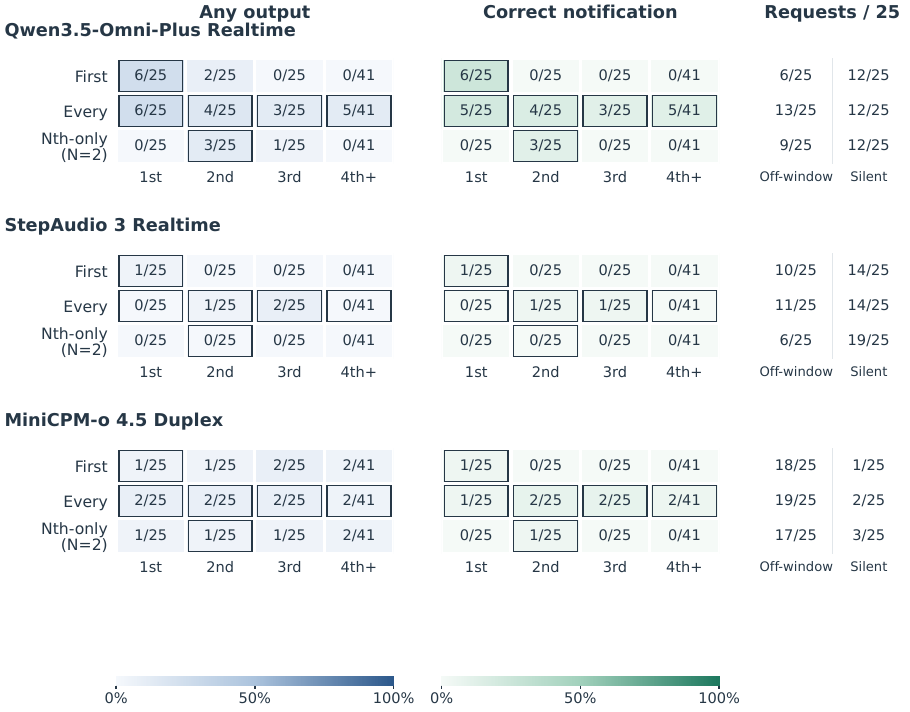}
\caption{Full rule contrasts for three native configurations, with the same counting conventions as Figure~\ref{fig:rule-appendix-simulated}. The timing window is $[-1,+3]$ seconds of audio progress. Early-event coverage should be read together with off-window outputs and silent requests.}
\label{fig:rule-appendix-native}
\end{figure}

\clearpage
\section{Auxiliary Fine-Tuning Experiment}
\label{app:auxiliary-finetuning}

We conduct an auxiliary sound-event fine-tuning experiment with Qwen3-Omni-30B-A3B-Instruct to examine how task-specific adaptation affects notification behavior with the same backbone. The training set contains 37,976 synthetic audio samples and 1,350,802 stepwise dialogue examples, including hard negatives in which the requested target is absent. We use LoRA~\citep{hu2022lora} with rank 8 and alpha 32, a learning rate of $2\times10^{-6}$, and a global batch size of 32. Training runs for 10,000 steps on 16 GPUs.

This experiment uses an auxiliary dataset and first-trigger protocol, separate from the 1,999-request benchmark. The base and fine-tuned models are evaluated on 1,112 queries constructed from 500 human-annotated audio samples, using the same audio, listening instructions, and trigger annotations. Table~\ref{tab:auxiliary-finetuning} reports this auxiliary first-trigger comparison.

\begin{table}[H]
\centering
\caption{Auxiliary sound-event first-trigger results (\%).}
\label{tab:auxiliary-finetuning}
\begingroup
\setlength{\tabcolsep}{12pt}
\renewcommand{\arraystretch}{1.08}
\begin{tabular}{lrr}
\toprule
Metric & Base model & Fine-tuned \\
\midrule
Precision & 51.21 & 67.72 \\
Recall & 30.52 & 26.39 \\
Negative-request FAR & 15.20 & 6.05 \\
\bottomrule
\end{tabular}
\endgroup
\end{table}

The fine-tuned model has higher precision and lower negative-request FAR, alongside reduced recall. This pattern is consistent with more conservative notification behavior and motivates further study of how task-specific training controls the recall--false-alarm trade-off. The experiment characterizes sound-event notification under this auxiliary protocol.

\end{document}